\documentclass[12pt, draftclsnofoot, onecolumn]{IEEEtran}
\usepackage{stmaryrd}
\usepackage{amsmath}
\usepackage{amssymb}
\usepackage{psfrag}
\usepackage{graphicx}
\usepackage{epstopdf}
\usepackage{todonotes}
\usepackage{cite}
\usepackage{algorithmic}
\usepackage{algorithm}
\usepackage{svg}

\usepackage{subfig}

\begin{document}
\title{{Learning for Detection: MIMO-OFDM Symbol Detection through Downlink Pilots}}
\author{\IEEEauthorblockN{Zhou Zhou, Lingjia Liu, and Hao-Hsuan Chang}\\
\IEEEauthorblockA{Bradley Department of Electrical and Computer Engineering, Virginia Tech., Blacksburg, VA, USA, 26041}}
\maketitle

\begin{abstract}
Reservoir computing (RC) is a special recurrent neural network which consists of a fixed high dimensional feature mapping and trained readout weights. In this paper, we introduce a new RC structure for multiple-input, multiple-output orthogonal frequency-division multiplexing (MIMO-OFDM) symbol detection, namely windowed echo state network (WESN). The theoretical analysis shows that adding buffers in input layers can bring an enhanced short-term memory (STM) to the underlying neural network. Furthermore, a unified training framework is developed for the WESN MIMO-OFDM symbol detector using both comb and scattered pilot patterns that are compatible with the structure adopted in 3GPP LTE/LTE-Advanced systems.
Complexity analysis suggests the advantages of WESN based symbol detector over state-of-the-art symbol detectors such as the linear minimum mean square error (LMMSE) detection and the sphere decoder, when the system is employed with a large number of OFDM sub-carriers. Numerical evaluations illustrate the advantage of the introduced WESN-based symbol detector and demonstrate that the improvement of STM can significantly improve symbol detection performance as well as effectively mitigate model mismatch effects compared to existing methods.
\end{abstract}

\section{Introduction}
Multiple-input, multiple-output, orthogonal frequency-division multiplexing (MIMO-OFDM) is the dominant wireless access technology for 4G and 5G cellular networks.
MIMO technology introduces additional spatial degrees of freedom and enables various multi-antenna transmission strategies such as transmit diversity, spatial multiplexing, and multi-user MIMO operations~\cite{LiuMIMOCom} to improve overall network performance.
To achieve the spatial-multiplexing gain in MIMO systems, different data streams are transmitted from different antennas causing inter-streams interference at the receiver. 
Accordingly, the symbol detection of multiple transmitted symbols from receiving antennas becomes critical for MIMO to realize its promise.
In general, MIMO detection is classified into coherent detection and non-coherent detection~\cite{yang2015fifty}.
In the coherent MIMO detection, the instantaneous channel matrix is obtained at the receiver through explicit channel estimation.
In this way, a two-step approach is adopted where the instantaneous channel matrix is estimated in the first step while MIMO symbol detection is conducted in the second step based on the estimated channel matrix as well as the received signals. On the other hand, in the non-coherent detection, the channel estimation is either performed implicitly or is completely avoided
where differential encoding is usually applied on input symbols leading to higher computational complexity.
Therefore, most modern wireless systems use coherent MIMO detection. 

OFDM technology combats the effect of frequency-selective fading by breaking a wide-band channel into multiple orthogonal flat-fading narrow-band channels to significantly simplify the transceiver architecture. 
However, the underlying time-domain waveform usually has a high peak-to-average power ratio (PAPR). 
This high PAPR reduce the power amplifiers (PAs) efficiency. Meanwhile, it produces input signal excursions into the PA's non-linear operation region resulting in signal distortions and spectral regrowth~\cite{rahmatallah2013peak}.
The non-linear distortion has a significant negative impact on the MIMO-OFDM channel estimation and symbol detection.
To address the non-linear distortion as well as the clipping noise, additional system resources are required to recover the distortion~\cite{chen2003iterative}. 
Alternatively, the digital pre-distortion (DPD), can be introduced ahead of the PA to compensate for PA's non-linearity effects~\cite{morgan2006generalized}. 
Note that a perfect knowledge of PA modeling and measurement bias is required for the DPD based compensation. However, obtaining this knowledge is very challenging in reality~\cite{joung2015survey}. 
Therefore, it is desirable to have a robust MIMO-OFDM symbol detector against the non-linear distortion. 

Artificial neural networks (NN) as an emerging technology provides new aspects for communication systems. For instance, an auto-encoder is introduced in~\cite{o2017introduction,S_Dorner} to conduct the symbol modulation. However, this end-to-end learning strategy often relies on a good channel model to facilitate the application.
\cite{farsad2018neural} employs recurrent neural network (RNN) as the receiver in molecular communication systems, where the underlying channel model is not available. Furthermore, in optical fiber systems \cite{karanov2018end,Karanov:19, khan2017machine}, ANNs are utilized as a channel equalizer as well as a network monitor. Especially, in \cite{karanov2018end}, the proposed method is verified through a lab experiment.

In light of the challenges in MIMO-OFDM symbol detection, NNs provide an ideal framework to conduct the symbol detection even under the non-linear distortion. 
In \cite{He2018}, a deep neural network (DNN) is introduced for OFDM symbol detection without using explicit channel state information (CSI). 
However, the offline training is conducted using available channel statistics. 
A NN-based method is introduced in \cite{zhang2019artificial} for the receiver design of a cyclic prefix (CP)-free OFDM system and a fully connected NN-based OFDM receiver is tested over the air in \cite{jiang2018artificial}. 
However, none of these works investigate MIMO-OFDM systems. 
In \cite{tan2018improving}, a DNN-based detector is introduced through unfolding the standard belief propagation algorithm. The parameters of the underlying DNN are required to be trained for different antenna configurations via an offline manner. In \cite{Samuel2018}, the feature of residual signals after layered processing is applied to construct a neural network for symbol detection. Meanwhile, the loss function is conducted on multiple layers in order to avoid the gradient vanishing \cite{szegedy2015going}. It demonstrates the introduced network can perform as well as the spherical decoding while achieving lower computational complexity. However, these methods require pre-known CSI as the coefficients or input of the underlying neural network which cannot be perfectly obtained when non-linear distortion exists. 

From the aforementioned examples, feedforward neural networks are employed for symbol detection by dividing the received signal into independent batches. 
On the other hand, communication signals are usually temporally correlated.  
Recurrent neural networks (RNNs) allow us to learn the temporal dynamic behaviors~\cite{goodfellow2016deep} making it a better tool for symbol detection. 
For standard RNN, the coefficients are often calculated via backpropagation through time (BPTT)~\cite{werbos1990backpropagation}.
However, when the sequence is inherent with long-range temporal dependencies, the training cannot converge due to the vanishing and exploding gradient, i.e., a small change at the current iteration can result in a very large deviation for later iterations\cite{pascanu2013difficulty}. To resolve the issue, RNNs are introduced with specific structures, such as the long short-term memory network \cite{hochreiter1997long} which uses “memory units” and “gating units” to control the gradient flow in order to avoid the gradient vanishing.
On the other hand, a large size of training set is required to construct a well-fitted RNN model.
However, the available training set for cellular networks especially in the physical layer is usually very limited due to the fact that the size of the training set is associated with the underlying system control overhead.
For example, in 3GPP LTE/LTE-Advanced systems, the pilot overhead is specified and is fixed for different MIMO configurations~\cite{LTE_standards}: The training set (demodulation reference signals) for SISO-OFDM is around 5\% of all the resource elements. 
On the other hand, for a $2 \times 2$ MIMO-OFDM system, the overhead for reference signals is around $10\%$. Therefore, how to effectively conduct RNN-based symbol detection for cellular networks under very limited training sets becomes important for realizing the promise of RNN in practical wireless networks.

Rooted in the backpropagation-decorrelation learning rule, reservoir computing (RC) is one type of RNNs, where the gradient issues of RNN training can be naturally avoided. 
More importantly, it can offer high computational efficiency with very limited training set~\cite{jaeger2001echo}. 
This is achieved by conducting learning only on the output layer where the untrained layers are sampled from a well-designed distribution. This makes RC an ideal tool for conducting symbol detection for cellular networks where the training set is extremely limited.
In fact, an RC-based MIMO-OFDM symbol detector is first introduced in our previous work~\cite{mosleh2017brain, mosleht2016energy}. 
With limited training set, \cite{mosleh2017brain} shows that the RC-based symbol detector can effectively combat the non-linear distortion caused by PA. 
However, our previous introduced RC-based symbol detector has limited performance using practical pilot patterns, such as the reference signal defined in LTE/LTE-Advanced standards. Since the wireless channel memory can introduce multi-path interference to the received signal, it motivates us to consider if an RC-based symbol detector with additional short term memory (STM) can improve the interference cancellation performance. Thus, the windowed echo state network (WESN) is introduced. The contributions of our paper are summarized as follows
\begin{itemize}
	\item We incorporated buffers\footnote{the buffer represents a linear shift register without any feedback tap} in the input layer of RC, i.e, WESN. Through theoretical analysis, we showed that the added buffer can improve the short-term memory of the underlying RC. Numerical evaluations also demonstrate a positive correlation between the detection performance and the improved short-term memory: WESN with improved short-term memory can perform better interference cancellation. A trade-off between the buffer length and the size of neurons is identified.
	
	\item We introduced a unified training method for WESN based on the pilot pattern which is compatible with the demodulation reference signal (DMRS) adopted in LTE/LTE-Advanced standards. In this way, we are able to demonstrate the fact that the introduced symbol detector can be effective under a very limited training set. To the best of our knowledge, this is the first work in the literature of conducting machine learning-based symbol detection using LTE/LTE-Advanced compatible pilot patterns. Meanwhile, we demonstrated the RC can detect symbols using non-orthogonal pilots through numerical evaluations.
	
	\item We analyzed the complexity of the RC-based symbol detector compared to conventional MIMO-OFDM receivers, such as linear minimum mean square error (LMMSE) and sphere decoding which is an approximation to the maximum likelihood estimator \cite{R2018,barbero2008fixing}. The results suggest that the RC-based detector has less computational complexity than conventional methods, especially when a large number of sub-carriers are utilized.
\end{itemize}

The structure of this paper is organized as follows: In Sec. \ref{System_Model_Intro_ESN}, the system model of MIMO-OFDM and conventional symbol detection methods are introduced. Meanwhile, the preliminary knowledge of reservoir computing is reviewed. In Sec. \ref{Symbol_detection}, the WESN based MIMO-OFDM symbol detector as well as the pilot structure are discussed. In addition, the analysis of the short term memory of WESN is presented in this section. The complexity comparison between conventional methods and the RC-based method is discussed in Sec. \ref{complex_analysis}. In Sec. \ref{PE}, the performance of WESN is evaluated. Finally, conclusions and future work are given in Sec.\ref{conclusion}.

\section{System Model and Preliminaries}
\label{System_Model_Intro_ESN}
\subsection{Channel Model and Transmitter Architecture}
We now consider the point-to-point MIMO-OFDM system, where the number of Tx and Rx antennas are respectively denoted as $N_t$ and $N_r$. 
At the $p$th transmitted antenna, the $i$th OFDM symbol is expressed as
\begin{align}
u^{(p)}_i(t) = \sum_{n = 0}^{N_c-1} x_{i}^{(p)}[n]  \exp({2\pi j n t/{ \Delta t}}), t \in [i \Delta t, (i+1)\Delta t),
\end{align}
where $x_i^{(p)}[n]$ is the transmitted symbol at the $n$th sub-carrier, $N_c$ stands for the number of sub-carriers, $\Delta t$ is the time length of one OFDM symbol\footnote{For simplicity, the index $t$ used in this paper can represent both analog and digital time index based on the context. When the t is related to digital processing components, an ADC is assumed as a prior to the processing. Otherwise, it represents the analog domain time index.}. At the $q$th antenna, the corresponding received OFDM symbol is given by
\begin{align}
\label{signal_model}
y_{i}^{(q)}(t) &= \sum_{p = 0}^{N_t - 1} h_{i}^{(q, p)}(t)\circledast g(u^{(p)}_i(t)) + n(t),
\end{align}
where $n(t)$ represents the additive noise, $\circledast$ stands for the circular convolution which is translated by the circular prefix of an OFDM symbol, $g(\cdot)$ is a general function of the waveform distortion which is discussed later in this section, and ${h^{(q, p)}_{i}(t)}$ is the channel response from the $p$th Tx antenna to the $q$th Rx antenna for the $i$th OFDM symbol.

Equivalently, the signal in Eq.~(\ref{signal_model}) can be rewritten in the digital frequency domain as
\begin{align}
\label{signal_model2}
{\tilde y}_{i}^{(q)}[n] = \sum_{p = 0}^{N_t - 1} {\tilde h}_{i}^{(p,q)}[n] {\tilde g}^{(p)}[n]  + {\tilde n}[n],
\end{align}
where ${\tilde n}[n]$ is the additive noise on the frequency domain, and
\begin{align}
{\tilde g}^{(p)}[n] = \int_{ \Delta t} g(u_i^{(p)}(t) )e^{-2\pi j t n/\Delta t}d t\\
{\tilde h}_{i}^{(p, q)}[n] = \int_{\Delta t} h_{i}^{(p,q)}( \tau) e^{- 2\pi n j \tau/{\Delta t}} d\tau.
\end{align}
When we set $g(z^{(p)}_i(t)) = z^{(p)}_i(t)$, we have 
\begin{align}
\label{observation_ideal}
{\tilde y}_{i}^{(q)}[n] &= \sum_{p = 0}^{N_t - 1}  {\tilde h}_{i}^{(p,q)}[n]x_i^{(p)}[n] +{\tilde n}[n].
\end{align}

\begin{figure}
	\centering
	\includegraphics[width=0.5\linewidth, height = 0.4\linewidth]{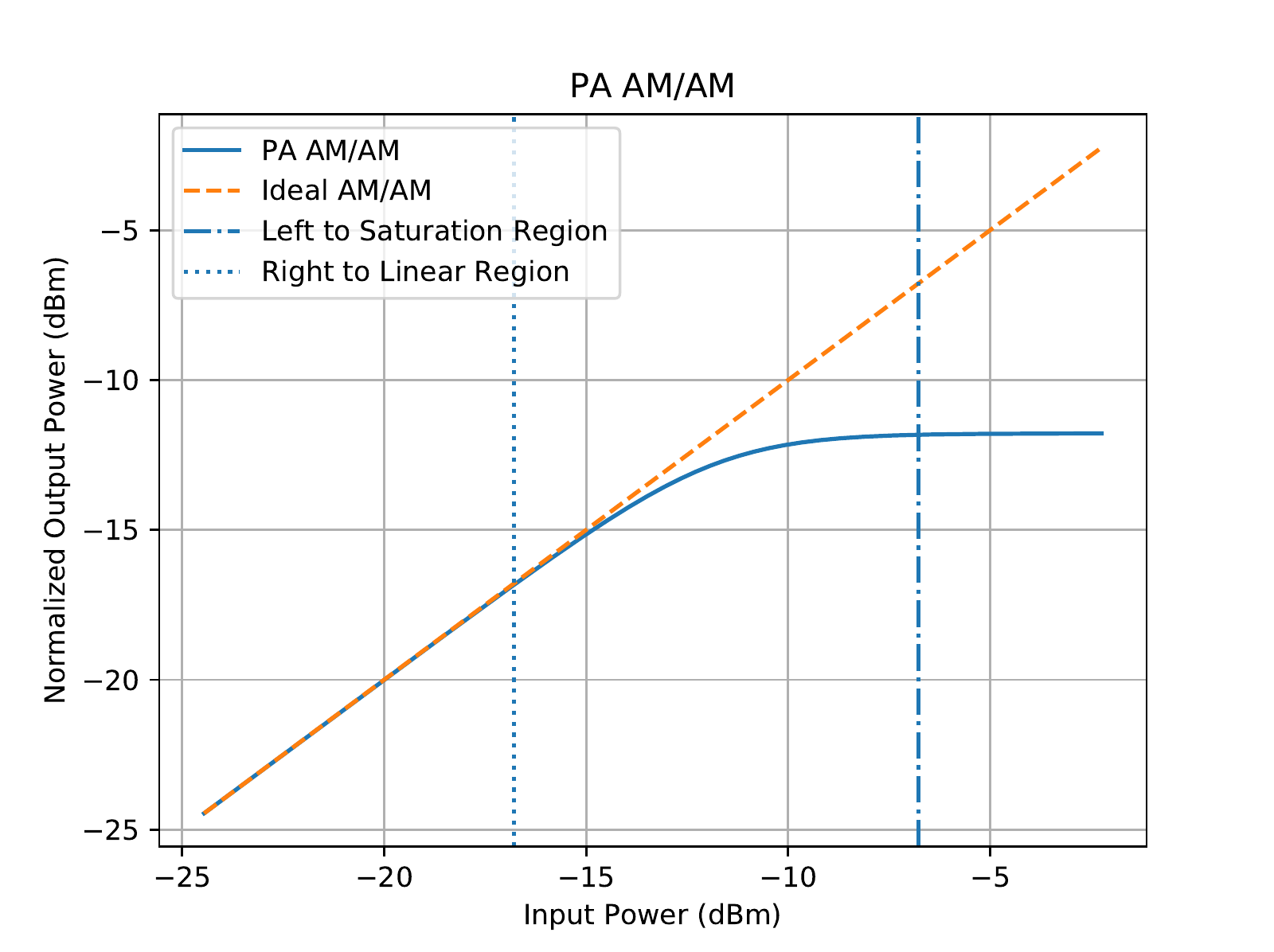}
	\vspace{-5mm}
	\caption{The input and output amplitude (AM/AM) curve of PA : $p = 3$ and $|u_{sat}|^2 = -11.78 {\text dB}$.}
	\label{PA_AM_AM}
	\vspace{-8mm}
\end{figure}

In the OFDM system, after the waveform is converted into the analog domain, it passes through RF circuits, such as power amplifiers, filters, and delay lines. 
These analog components are usually nonlinear systems due to practical constraints (e.g., circuit spaces and power consumption). 
For instance, the input-output relation of the power amplifier (PA) can be represented using the RAPP model\cite{joung2015survey}:
\begin{align}
g(u(t)) = {{G_0u(t)}\over {\left[1+\left({{|u(t)|}\over {u_{sat}}}\right)^{2p}\right]^{1/{2p}}}}
\end{align}
where $u(t)$ is the input signal of PA, $G_0$ stands for the power gain of PA, $u_{sat}$ is the saturation level, and $p> 0 $ is the smooth factor. 
The corresponding operational region of the PA is shown in Fig. \ref{PA_AM_AM} where the region is generally divided into three parts: the linear region $|u(t)|\ll u_{sat}$, the non-linear region $|u(t)|\sim u_{sat}$, and the saturation region $|u(t)|\gg u_{sat}$. 
Even though the signal waveform is perfectly retained in the linear region, the power efficiency is low. 
Therefore, to reduce the distortion while maintaining relatively high efficiency, the PA operational point is set in the linear region that is close to the nonlinear region. 
Meanwhile, due to the high peak average power ratio (PAPR) of the OFDM signal, PAPR reduction is also employed to guarantee a certain level of PA efficiency \cite{rahmatallah2013peak}. 
However, the consideration of PA efficiency will lead to the deficiency in transmission reliability due to the underlying waveform distortion. 
In this paper, we denote the resulting distortion as a function $g(\cdot)$. 

\subsection{Conventional Methods}
\label{Conventional_Detection}
\begin{figure*}[!t]
	\centering 
	\subfloat[]{\includegraphics[width=0.6\linewidth, height = 0.25\linewidth]{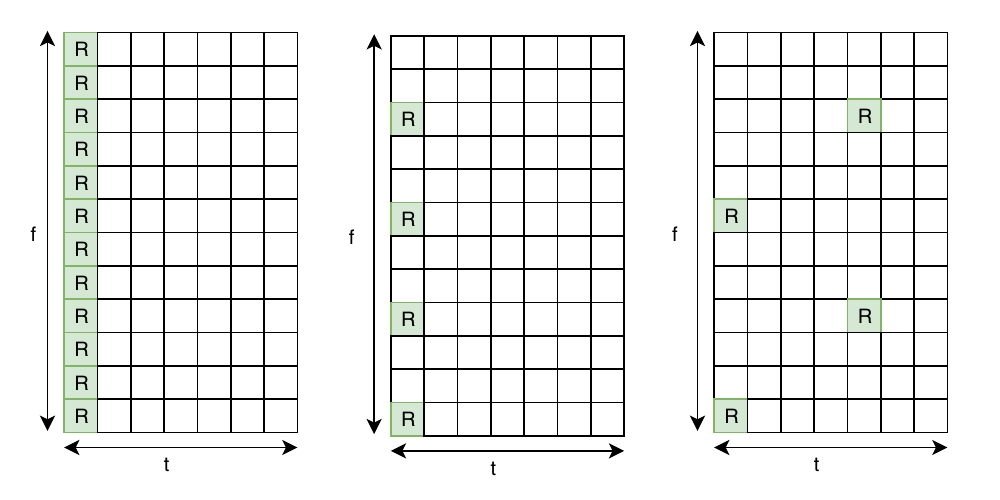}
		\label{fig_OFDM_pilots1}}
	\vfil
	\subfloat[]{\includegraphics[width=0.7\linewidth, height = 0.25\linewidth]{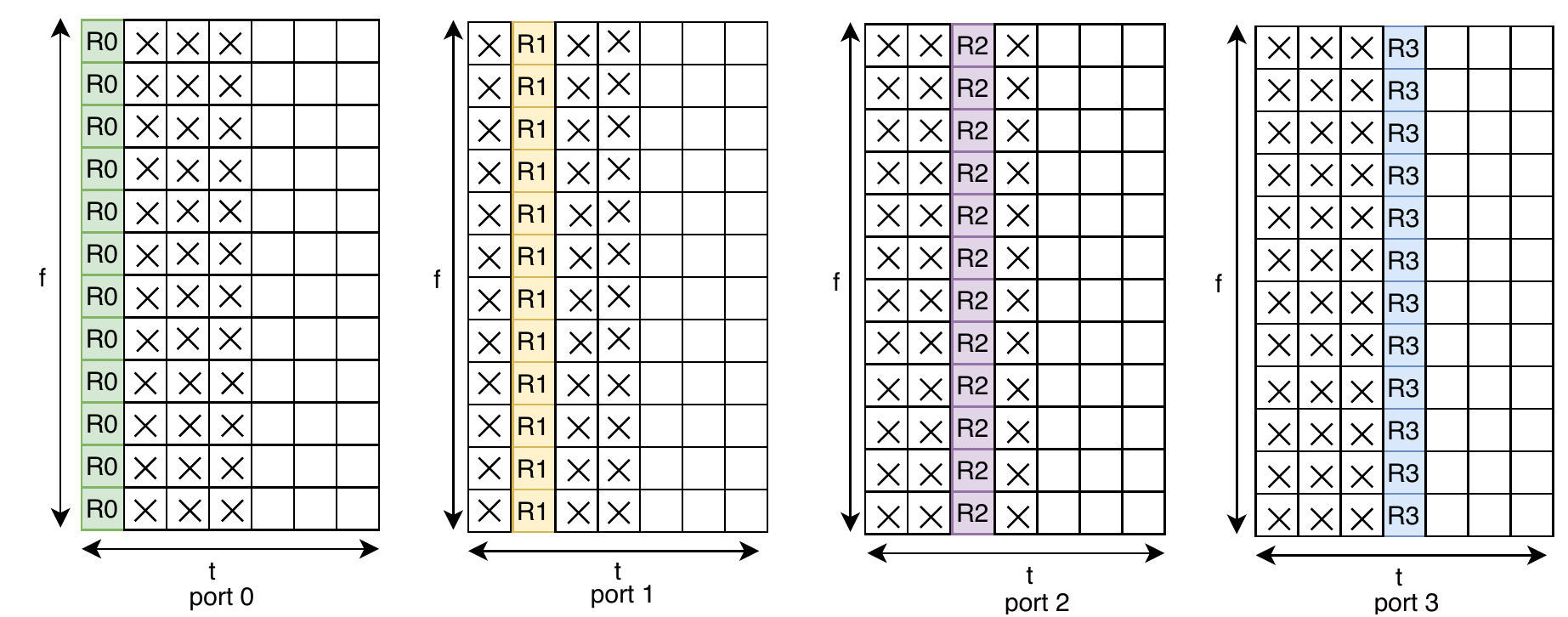}
		\label{fig_OFDM_pilots2}}
	\vfil
	\subfloat[]{\includegraphics[width=0.7\linewidth, height = 0.25\linewidth]{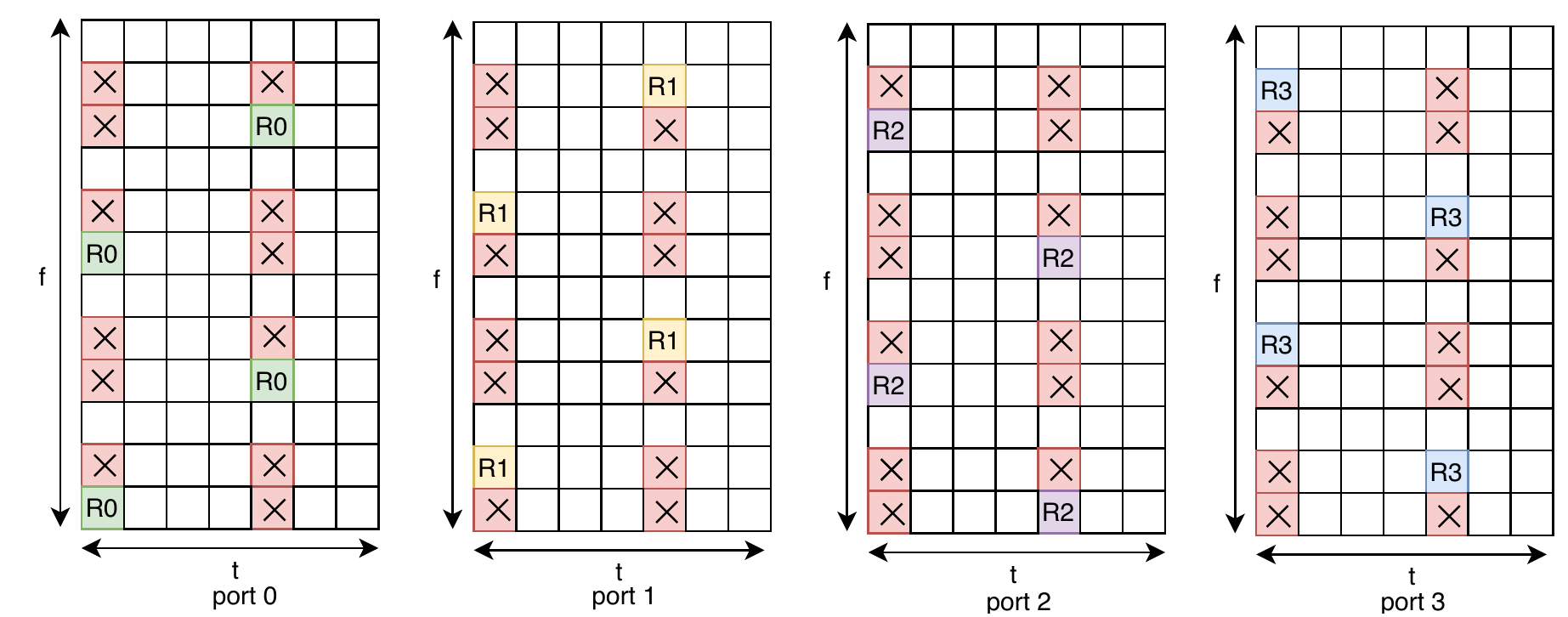}
		\label{fig_OFDM_pilots3}}
	\caption{OFDM pilots structures in one RB (a) SISO-OFDM pilots (b) comb structured MIMO-OFDM pilots  (c) scattered structured MIMO-OFDM pilots.  } 
	\label{fig_OFDM_pilots}
	\vspace{-8mm}
\end{figure*}

Coherent symbol detection methods are conducted by a two steps: channel estimation and symbol detection. In the channel estimation, a series of pre-known pilots ${\bar x}_{i}^p[n]$ is sent to Rx, where ${i \in \Omega_t}$, ${p \in \Omega_s}$, $n \in \Omega_f$ in which $\Omega_t$, $\Omega_s$ and $\Omega_f$ respectively represent the pilot index sets of OFDM symbols, antennas, and sub-carriers. 
Specifically, in LTE/LTE-Advanced systems, the design of pilot patterns is based on resource blocks (RB) as shown in Fig. \ref{fig_OFDM_pilots}. For single input single output (SISO) OFDM systems, the pilot structures are depicted in Fig. \ref{fig_OFDM_pilots} (a). The first sub-figure illustrates that $\Omega_t$ equals to the first OFDM symbol and $\Omega_f$ occupies all the sub-carriers. 
This comb pattern can be applied to the block fading channel assumption which is used in~\cite{mosleh2017brain}. 
The size of $\Omega_f$ can be further reduced as shown in the second sub-figure of Fig. \ref{fig_OFDM_pilots} (a) where the channel interpolation can be incorporated using frequency coherence. 
In the third subfigure of Fig. \ref{fig_OFDM_pilots} (a), the scattered pilot pattern is applied on a Doppler channel which facilitates the channel tracking with very limited pilot overhead. 
For the MIMO channel, the pilot pattern is shown in Fig. \ref{fig_OFDM_pilots} (b) and \ref{fig_OFDM_pilots} (c). 
In Fig. \ref{fig_OFDM_pilots} (b), the pilot symbols at different antenna ports are non-overlapping since they are allocated to different OFDM symbols. 
In Fig. \ref{fig_OFDM_pilots} (c), the cross marker represents the null pilot symbols. 
Therefore, the pilot interference is eliminated during the channel estimation stage for MIMO. 

Using pilots, the channel coefficients on the corresponding resource elements (REs) are obtained through (\ref{signal_model2}) by solving
\begin{align}
\label{CSI_estimation}
\min_{{\tilde h}^{(p, q)}_i[n]} l({\tilde { y}}_{i}^{(q)}[n] , {\bar x}_i^{(p)}[n]|(i,p,n)\in \Omega_t\times\Omega_s\times \Omega_f)
\end{align}
where $l(\cdot)$ is a pre-defined loss-function, such as likelihood function, mean square error, etc.. 
The CSI on the rest of the REs is inferred through an interpolation method. By substituting the estimated ${\hat { h}}_{i}^{(p, q)}[ n]$ into (\ref{signal_model2}), the rest symbols $\{x_{i}^{(p)}[n]|(i,p,n)\in (\Omega_t\times\Omega_s\times \Omega_f)^c \}$ (where $\Omega^c$ stands for the complementary set of $\Omega$) are estimated using
\begin{align}
\label{symbol_detection}
\min_{{ x}^p_i[n]} l({\tilde { y}}_{i}^{(q)}[n] , {\hat h}_i^{(p, q)}[n]|(i,p,n)\in (\Omega_t\times\Omega_s\times \Omega_f)^c).
\end{align}
However, the optimal solutions for (\ref{CSI_estimation}) and (\ref{symbol_detection}) are not usually guaranteed due to the nonlinear distortion $g(\cdot)$. An improper assumption on $g(\cdot)$ can cause the model mismatch which deteriorates the accuracy on solving the estimation problem (\ref{CSI_estimation}) and the detection problem (\ref{symbol_detection}). 
To circumvent this dilemma, i.e., the dependence on the model assumption, RC based method can be employed as an alternative solution.
\subsection{Reservoir Computing}
\label{Reservoir_computing_intro}
\begin{figure}
	\centering
	\includegraphics[width=0.55\linewidth, height = 0.35\linewidth]{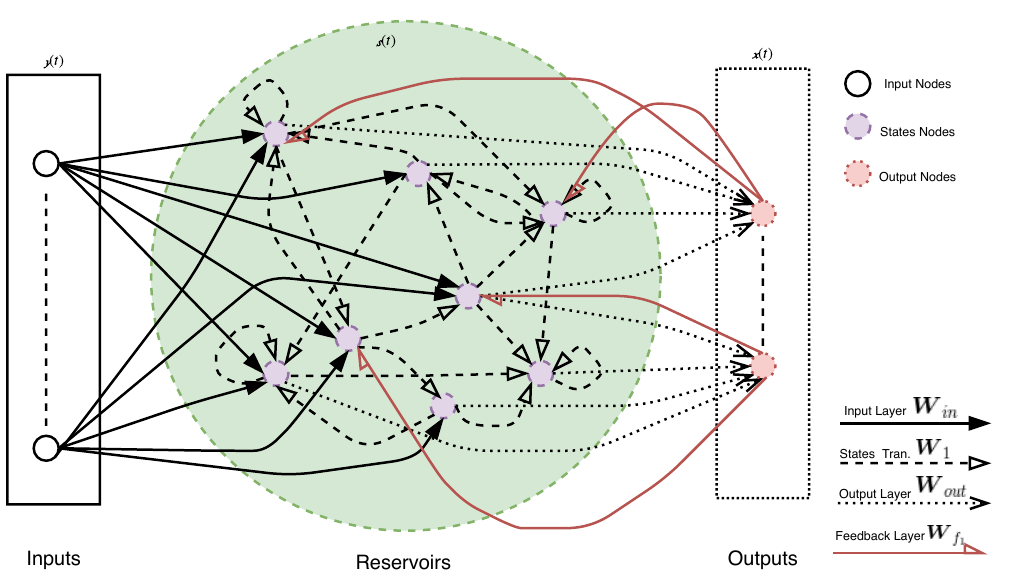}
	\vspace{-5mm}
	\caption{An example of reservoir computing, the echo state network.}
	\label{fig:pic-2}
	\vspace{-8 mm}
\end{figure}

In this section, we will briefly introduce the basic structure of RC. 
RC is one category of RNNs which consists of an input mapping, a fixed dynamic system, and a trained readout network. 
In general, there are two types of RC network architectures: echo state network (ESN) and liquid state machine (LSM).
The network architecture of the ESN~\cite{jaeger2001echo} is illustrated in Fig.~\ref{fig:pic-2} where the underlying network dynamics can be described by the following equation
\begin{align}
\label{state_equation}
{\boldsymbol s}(t+1) = f_{states}({\boldsymbol W}'[{\boldsymbol y}^T(t+1), {\boldsymbol s}^T(t), {\boldsymbol x}^T(t)]^{T})
\end{align}
where ${\boldsymbol s}(t)\in {\mathbb C}^{N_n}$ represents the inner states, $N_n$ is the number of neurons inside the reservoir, ${\boldsymbol y}(t)$ is the input signal, $f_{states}$ represents the states activation function, ${\boldsymbol W}' = [{\boldsymbol W}_{in}, {\boldsymbol W}_{1}, {\boldsymbol W}_{f_1}]$, where ${\boldsymbol W}_{in}$ is weights of the input layer, ${\boldsymbol W}_{1}$ is the inner state transition weights, and ${\boldsymbol W}_{f1}$ is weights of the feedback layer. Moreover, ${\boldsymbol W}_{f1}$ can be omitted when feedback is not required. The output equation is given by
\begin{align}
\label{output_equation}
{\boldsymbol x}(t+1) = f_{out}({\boldsymbol W}_{out} {\boldsymbol s}^T(t+1))
\end{align}
where $f_{out}$ is the activation function, and ${\boldsymbol W}_{out}$ represents the output layer. ${\boldsymbol W}'$ is designed according to the following {\it echo state property}.
\newtheorem{definition}{Definition}
\newtheorem{theorem}{Theorem}
\begin{definition}
	We consider an ESN following the state transition equation (\ref{state_equation}). Given an input sequence ${\boldsymbol y}(t)$ and two finite initial states ${\boldsymbol s}_1(0)$ and ${\boldsymbol s}_2(0)$, for any $\epsilon >0$ and ${\boldsymbol y}(t)$, if we have $\|{\boldsymbol s}_1(t) - {\boldsymbol s}_2(t)\|<  \epsilon$ when $t > \kappa(\epsilon)$, where $\xi(\epsilon)$ is a positive number, then the ESN satisfies the echo state property.
\end{definition}
Nevertheless, the echo state property of a given ESN cannot be easily justified from the above definition. For ease of application, the following sufficient condition is usually applied.
\begin{theorem}[Proposition 3 in \cite{jaeger2001echo}]
	Assume an ESN with $\tanh(\cdot)$ as the activation function. If the maximum singular value of the inner states transition weight matrix $\boldsymbol W$ is smaller than 1, i.e., $\sigma({\boldsymbol W})_{max} < 1$, then for all input $\boldsymbol y(t)$ and initial states ${\boldsymbol s} \in [-1, 1]^N $, the ESN satisfies the echo state property.
\end{theorem}

Learning of output weights ${\boldsymbol W}_{out}$ contains the following two stages:
\begin{itemize}
	\item Generation of the states trajectory: By feeding the training input $\{{\boldsymbol {\bar y}}(t)\}_{t = 0}^T$ into ESN with target $\{{\boldsymbol {\bar x}}(t)\}_{t = 0}^T$, the states set $\{{\boldsymbol {\bar s}}(t)\}_{t = 0}^T$ is obtained by (\ref{state_equation}), where $T$ represents the sequence length of the training input.
	\item Regression on the output weights: Substituting the generated states $\{{\boldsymbol {\bar s}}(t)\}_{t = 0}^T$ into (\ref{output_equation}), we can calculate the weights ${\boldsymbol W}_{out}$ through 
	\begin{align}
	\min_{{\boldsymbol W}_{out}}  L(\{{\boldsymbol {\bar y}}(t)\}_{t = 0}^T, \{f_{out}({\boldsymbol W}_{out}{\boldsymbol {\bar s}}^T(t))\}_{t = 0}^T).
	\end{align}
\end{itemize}
Specifically, when we choose $f_{out}$ as an identity function, $L$ as Frobenius norm, the output weights are solved by
\begin{align}
\label{learning_rule}
\min_{{\boldsymbol W}_{out}} \sum_{t = 0}^T\|{\boldsymbol {\bar y}}(t) - {\boldsymbol W}_{out}{\boldsymbol {\bar s}}(t)\|_F^2
\end{align}
which has a closed-form solution as follows
\begin{align}
{\boldsymbol W}_{out} ={\bar{\boldsymbol Y}}{\boldsymbol {\bar S}}^+,
\end{align}
where $\bar{\boldsymbol Y} = [{\boldsymbol {\bar y}}(0),\cdots, {\boldsymbol {\bar y}}(T)]$, ${\boldsymbol {\bar S}} = [ {\boldsymbol {\bar s}}^T(0),\cdots, {\boldsymbol {\bar s}}^T(T)]$, and ${\boldsymbol {\bar S}}^+$ is the Moore-Penrose inverse of $\boldsymbol {\bar S}$.

\section{Symbol Detection}
\label{Symbol_detection}
\subsection{Neural Network Based Approach}


The neural network based symbol detection consists of two steps: training and testing. In the training stage, base station (BS) sends pre-defined symbols $\{{\bar x}_{i}^{(p)}[n]| (i,p,n)\in \Omega_t\times\Omega_s\times \Omega_f\}$ to mobile stations (MSs). Then, MSs train a neural network receiver $\mathcal D$ by solving
\begin{align}
\label{Learning_NN}
\min_{\mathcal D} f({\mathcal D}({\bar y}_{i}^{(p)}(t)), {\bar x}_{i}^{(p)}[n]| (i,p,n) \in \Omega_t\times\Omega_s\times \Omega_f),
\end{align}
where $f(\cdot)$ is the training objective function and $\mathcal D$ is the neural network; ${\bar y}_{i}^{(p)}(t)$ represents the received signal at a MS in the training stage. For instance, $f(\cdot)$ can be the mean squared error or the cross-entropy; $\mathcal D$ can be the fully connected, convolution or recurrent neural networks. 
In the testing stage, the symbols are estimated by feeding the observation ${ y}_{i}^{(p)}(t)$ to the learned neural network $\mathcal {\hat D}$, i.e., ${\mathcal {\hat D}}(y_{i}^{(p)}(t))$. Consequently, the symbol detection performance and implementation complexity are determined by the utilized neural network and learning method. However, in wireless communications, the resources allocated to pilots are much less than the transmitted data symbols. 
Therefore, overfitting can occur if the adopted NN structure is not carefully designed. 
\subsection{Windowed Echo State Network}
\label{WESN}
\subsubsection{ESN Short Term Memory}
For RNN, the output features are expected as a certain function of the memory encoded from inputs. A longer memory allows wider time-spanned features to be learned. Intuitively, the memory size can be characterized as the ability of recovering historical inputs. Thus, the memory capacity of ESN is defined as follows:
\begin{definition}[Short Term Memory \cite{jaeger2001short}]
	Given an ESN with fixed coefficients of the inner state transient matrix, input layer, and activation function, we first define the following {\it self-delay reconstruction correlation} 
	\begin{align}
	d(m, {\boldsymbol w}_{out}) ={cov(y(n - m), x(n)) \over {\sigma(y(n-m)) \sigma(x(n))}},
	\end{align}
	where ${\boldsymbol w}_{out}$ is the output weight for the ESN with a single input and a single output; With a slight abuse of notations, in this subsection, $n$ represent the time sequence index, $m$ is the input delay degree, and $x(n)$ is the ESN output when input is $y(n)$. Then, relying on the self-delay reconstruction correlation, we have the following definitions,
	\begin{itemize}
		\item 	The $m$-th delay STM capacity:
		\begin{align}
		\label{MC_i}
		MC_m = \max_{{\boldsymbol w}_{out}} d(m,{{\boldsymbol w}_{out}}).
		\end{align}
		\item The STM capacity:
		\begin{align}
		MC = \sum_{ m = 1,2,\cdots } MC_m.
		\end{align}
	\end{itemize}
\end{definition}
Remark that the above definition is only for ESN with a single input and a single output. The general definition of STM for ESN with multiple inputs and multiple outputs is obtained by extending the concept to each input-output pair. Furthermore, the metric in (\ref{MC_i}) can be approximately calculated through a self-delay training procedure defined as follows: 1) Input the zero mean sequence $\{y(n)\}_{n = 0}^{N-1}$ to ESN; 2) Train the output $\{x(n)\}_{n=m}^{N-1}$ using the target $\{y(n)\}_{n = 0}^{N-m-1}$, where 
$x(n) ={\boldsymbol w}_{out} {\boldsymbol s}(n)$ and ${\boldsymbol{s}}(n)$ is the state of the ESN. Therefore, the {\it self-delay reconstruction correlation} can be rewritten as 
\begin{align}
d(m, {\boldsymbol w}_{out}) = {\sum_{n = 0}^{N - m - 1} y(n)x(n + m)\over {\sqrt{\sum_{n = 0}^{N - m - 1} |y(n)|^2}  \sqrt{\sum_{n = m}^{N-1} |x(n)|^2}}} \\
\propto -\|\tilde{\boldsymbol { x}}(m:N-1) - \tilde{\boldsymbol y}(0:N-m-1)\|_2^2,
\end{align}
where $\propto$ stands for in a relation of proportionality; $\tilde{\boldsymbol { x}}(m:N-1)$ is a normalized vector stacked by the samples from $x(m)$ to $x(N-1)$; and $\tilde{\boldsymbol y}(0:N-m-1)$ is stacked by samples from $y(0)$ to $y(N-m-1)$. According to the output equation of ESN in (\ref{output_equation}), $\tilde{\boldsymbol { x}}(m:N-1)$ can be equivalently expressed as
\begin{align} 
\label{weight_learning}
\tilde{\boldsymbol { x}}(m:N-1) = {{\boldsymbol w}}_{out}{\tilde {\boldsymbol S}},
\end{align}
where ${\tilde{\boldsymbol S}} = [ {\tilde{\boldsymbol s}}^T(m), {\tilde{\boldsymbol s}}^T(m + 1),\cdots, {\tilde{\boldsymbol s}}^T(N - 1)]$ and $\tilde{\boldsymbol s}(n)$ denotes the scaled states such that ${\tilde {\boldsymbol s}}(n) = {\boldsymbol s}(n)/\|{\boldsymbol w}_{out}{\boldsymbol s}(n)\|_2$. From the above definition, we can obtain the STM capacity of the buffer as follows
\begin{theorem}
	\label{theorem1}
	The memory capacity of a buffer is greater than $M$, where $M$ is the buffer's size.
\end{theorem}

\begin{IEEEproof}
	For a buffer, it is easily known that $MC_m = 1$, if $0 \leq m \le M$. When $m > M$, we have $MC_{m} \geq 0$ as the signal can be self-correlated. Therefore, we have $MC_{W} \geq M$.
\end{IEEEproof}
Furthermore, we have the following upper bound for the STM capacity of ESN
\begin{theorem}[Proposition 2 in \cite{jaeger2001short}]
	\label{theorem0}
	The memory capacity of ESN is bounded by the number of neurons, i.e., $MC_{ESN}< N_n$. 
\end{theorem}

\begin{figure*} 
	\centering
	\includegraphics[width=0.85\linewidth, height = 0.35\linewidth]{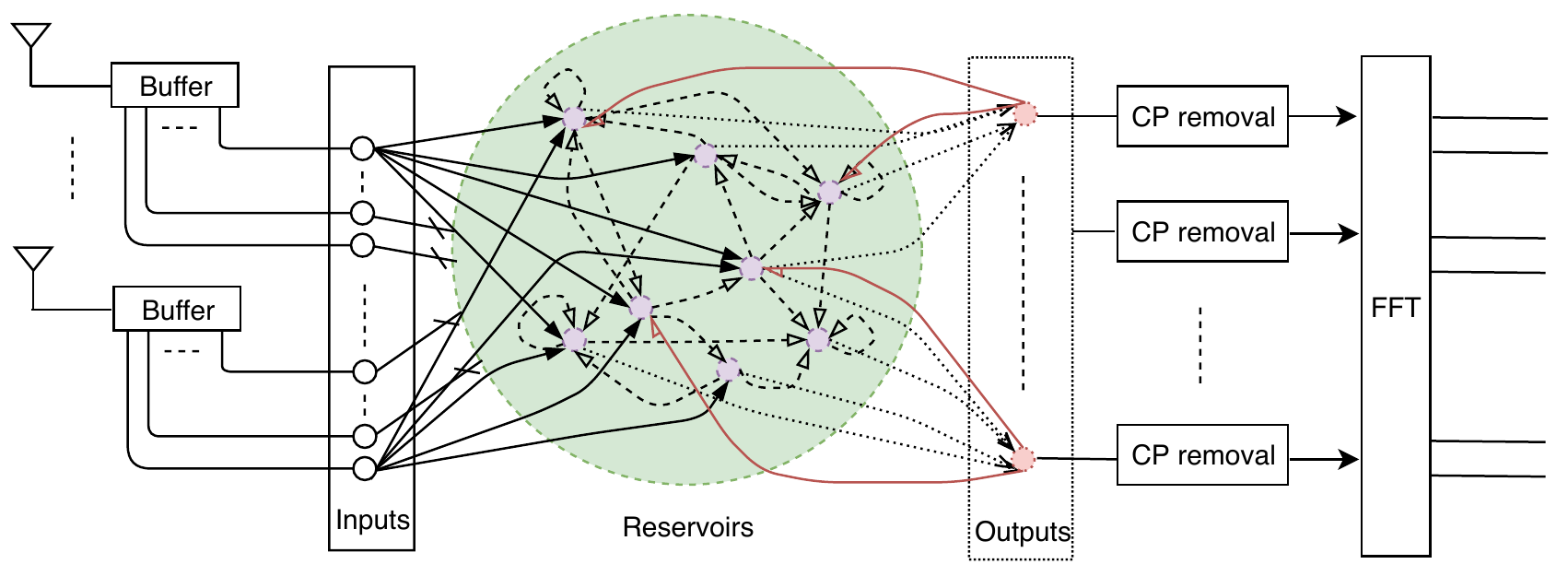}
	\vspace{-5mm}
	\caption{The architecture of WESN based MIMO-OFDM symbol detector.}
	\label{RC_Rx_Arch}
	\vspace{-8 mm}
\end{figure*}

Note that the above conclusion can only be made when the network is with an identity output activation and an i.i.d input. However, this theorem can give us a general guide on setting the number of neurons. Comparing Theorem \ref{theorem1} to Theorem \ref{theorem0}, we see the buffer has a higher STM capacity than ESN when the buffer size is the same as the number of neurons of the ESN. However, a higher STM capacity does not necessary stand for a better nonlinear feature mapping ability. This is because reservoirs process the input history through a highly nonlinear recursive procedure rather than simply preserve the input. In our extension, adding a buffer at the input of ESN as depicted in Fig. \ref{RC_Rx_Arch}, we can obtain the WESN. Its STM is characterized by
\begin{theorem}
	\label{theorem2}
	Given a WESN, suppose the STM capacity of the buffer and the ESN component are $MC_W$ and $MC$ respectively. Then, the STM capacity of the WESN, $MC_{WESN}$, is given by
	\begin{align}
	{1\over 2}MC_{WESN} \geq \lambda MC_{W} + (1-\lambda)MC_{ESN}, \lambda \in (0, 1).
	\end{align}
\end{theorem}
\begin{IEEEproof}
	See Appendix \ref{Meomery_ESN}.2 for details.
\end{IEEEproof}
The above result shows that WESN can achieve a higher STM capacity than the convex combination of the buffer STM and ESN STM. 

\subsection{WESN based MIMO-OFDM receiver}
The introduced WESN based MIMO-OFDM symbol detector is shown in Fig. \ref{RC_Rx_Arch}. 
We see the receiving link is concatenated by a WESN, a cyclic prefix (CP) removal and an FFT block, where the dimension of the WESN outputs is the same as the number of the transmission streams. The received $i$th OFDM symbol ${\boldsymbol y}_{i}(t) =  [y^{(0)}_{i}(t), y^{(1)}_{i}(t),\cdots, y^{(N_r - 1)}_{i}(t)]^T$ is first fed into the buffers. At the $j$th antenna's buffer, it collects $N_{bf}$ samples from $y_i^{(j)}(t)$ to create a vector $[y_{i}^{(j)}(t-N_{bf}), y_{i}^{(j)}(t-N_{bf}+1),\cdots, y_{i}^{(j)}(t)]^T$. Thereafter, the vector is mapped into reservoirs through the input-layers. Reservoirs update their inner states and generate an output vector ${\boldsymbol z}_{i}(t) = [z^{(0)}_{i}(t), z^{(1)}_{i}(t),\cdots, z^{(N_r - 1)}_{i}(t)]^T$, where ${\boldsymbol z}_{i}(t) \in {\mathcal C}^{N_r}$ and ${\mathcal C}$ represents the modulation constellation. Finally, it converts ${\boldsymbol Z}_{i}$ into the frequency domain, where ${\boldsymbol Z}_{i} = [{\boldsymbol z}_{i}(0), {\boldsymbol z}_{i}(1), \cdots, {\boldsymbol z}_{i}(N_c-1)]\in{\mathbb C}^{N_r \times N_c}$, and quantize the resulting frequency signal into modulation symbols according to the constellation $\mathcal C$, i.e., ${\mathcal Q}_{\mathcal C}({\boldsymbol Z}_{i} {\boldsymbol F})$, where $\boldsymbol F$ represents the Fourier transform matrix.
\subsection{Training of WESN}
We begin by considering the training the WESN receiver under the SISO channel with zero Doppler shift, i.e., $f_D = 0$. As discussed in Sec. \ref{Conventional_Detection}, we assume the first OFDM symbol is the training set. According to (\ref{Learning_NN}), we select the objective function $f$ as the Frobenius norm induced distance and $\mathcal D$ as the WESN. Using the ESN's dynamics and output equations in Sec. \ref{Reservoir_computing_intro}, we have the output of WESN as $ {\boldsymbol W}{\boldsymbol S}$, where ${\boldsymbol S}\in {\mathbb C}^{(N_n) \times N_c}$ stands for the reservoir states, and ${\boldsymbol W}\in {\mathbb C}^{1 \times {(1+N_n)}}$ is the readout weights. With a slight generalization in our notations, here $N_n$ stands for the number of neurons plus the lenght of buffers.
Therefore, similarly as (\ref{learning_rule}), the readout weights of the WESN are updated by solving
\begin{align}
\label{SISO_output_learning}
\min_{{\boldsymbol W}} \|{\boldsymbol W}{\boldsymbol S}{\boldsymbol F} - {\boldsymbol {\bar x}}^T_0\|_2,
\end{align}
where ${\boldsymbol F}\in {\mathbb C}^{N_c \times N_c}$ represents the Fourier transform matrix, and ${\boldsymbol {\bar x}}_0\in {\mathcal C}^{N_c}$ is the pilot symbols in which the subscript stands for the first OFDM symbol. The solution can be further written as the following closed-form,
\begin{align}
\label{ESN_output_weights}
{\boldsymbol W}{ \stackrel {(a)}{=}} {\boldsymbol {\bar x}}^T_0 ({\boldsymbol S}{\boldsymbol F})^+ \stackrel {(b)}{=} ({\boldsymbol {\bar x}}^T_0{\boldsymbol F}^H) {\boldsymbol S}^+,
\end{align}
where $(a)$ holds when we assume the number of training symbols is greater than the number of neurons plus inputs. Alternatively, through $(b)$, the weights learning can be interpreted as fitting the output of WESN to the waveform of the target OFDM symbols ${\boldsymbol {\bar x}}^T_0{\boldsymbol F}^H$. 

\begin{figure*}[!t]
	\centering 
	\subfloat[]{\includegraphics[width=0.7\linewidth, height = 0.25\linewidth]{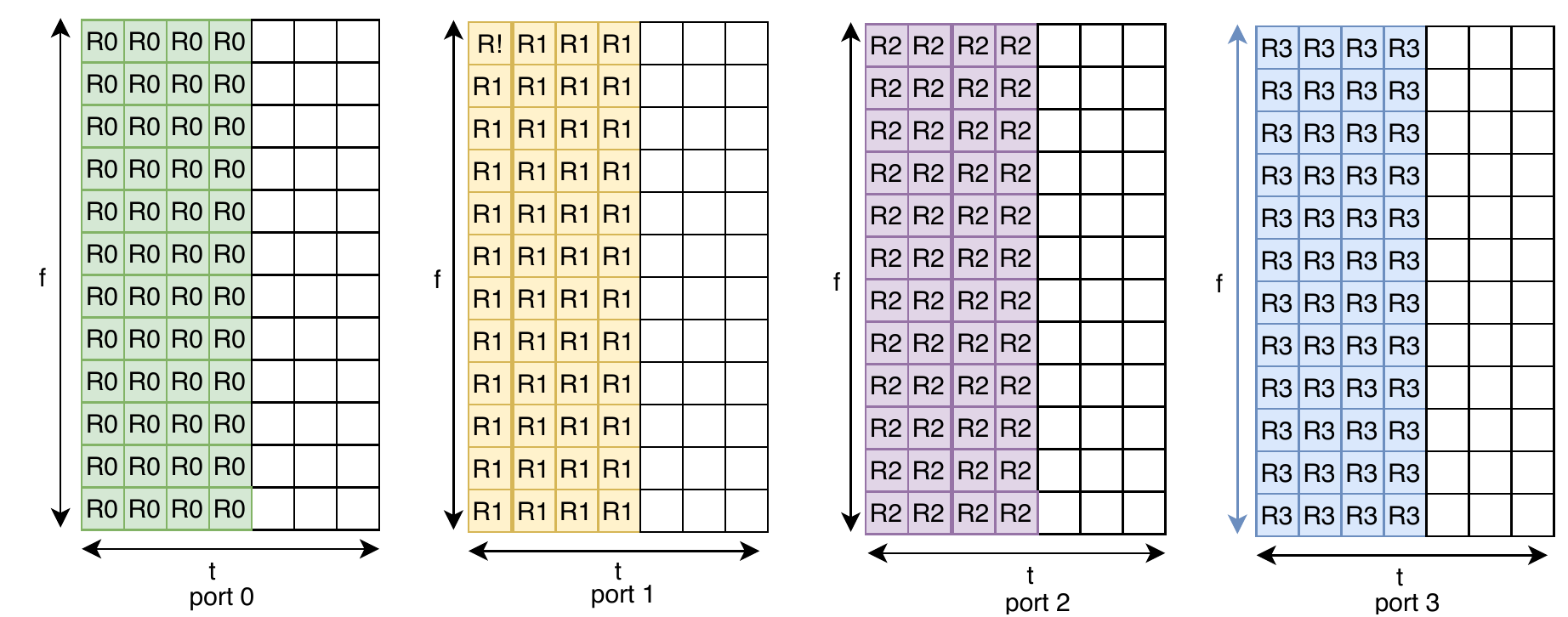}
		\label{fig_ESN_MIMO_pilots1}}
	\vfil
	\subfloat[]{\includegraphics[width=0.7\linewidth, height = 0.25\linewidth]{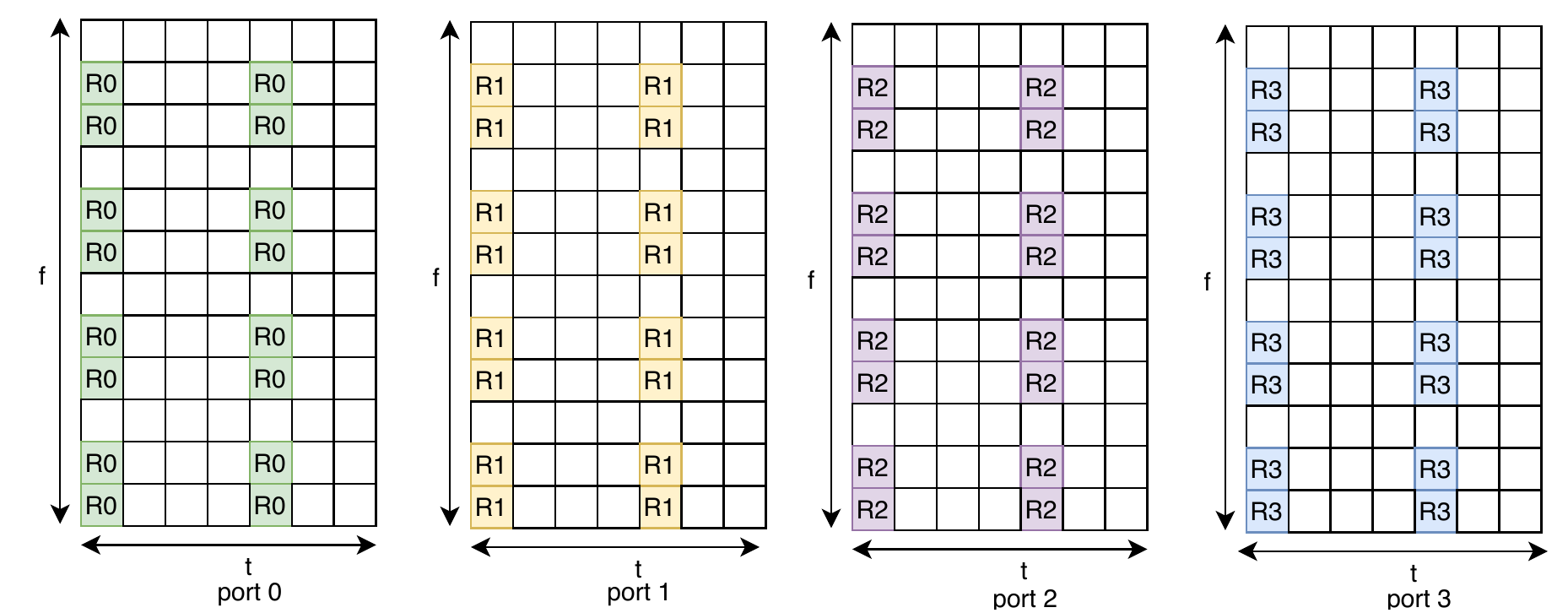}
		\label{fig_ESN_MIMO_pilots2}}
	\caption{The OFDM pilots structures for WESN in one RB: (a) block (b) scattered. } 
	\label{fig_ESN_MIMO_pilots}
	\vspace{-8 mm}
\end{figure*}

We then extend the symbol detection method to the MIMO channel with a zero Doppler shift. Rather than SISO, the MIMO receiver needs to mitigate the inter-streams interference. 
To realize this, a tailored training pilot pattern is introduced, where Fig. \ref{fig_ESN_MIMO_pilots1} shows the case of $N_t = N_r =  4$. 
This pattern occupies the same number of REs as the comb structured MIMO-OFDM pilots in Fig. \ref{fig_OFDM_pilots2}. 
There is a slight difference between these two patterns: in Fig. \ref{fig_OFDM_pilots2}, the pilots from different antennas are orthogonal to each other, while those in Fig. \ref{fig_ESN_MIMO_pilots1} are overlapping. 
This is due to the fundamental difference between learning-based methods and conventional channel estimation-based methods: In learning-based methods, the neural networks need to learn the interference situation of the transmission; in conventional methods, received pilots should be interference-free to improve channel estimation performance.

By using this pilot pattern, the outputs of the WESN can be expressed as the matrix ${\boldsymbol { Z}} = [{\boldsymbol { Z}}_{ 0},{\boldsymbol { Z}}_{ 1},{\boldsymbol { Z}}_{ 2},{\boldsymbol { Z}}_{3}]$, where the subscripts represent the indices of the OFDM symbols allocated as pilots. Similarly, we have ${\boldsymbol { Z}} = {\boldsymbol W}{\boldsymbol S}$, where ${\boldsymbol S} = [{\boldsymbol S}_{0},{\boldsymbol S}_{1},{\boldsymbol S}_{ 2}, {\boldsymbol S}_{3}]\in {\mathbb C}^{N_n \times 4N_c}$ represents the state matrix of WESN. Thus, the output layer is solved by
\begin{align}
\min_{{\boldsymbol W}} \|{\boldsymbol W}{\boldsymbol S}{\boldsymbol F}' - {\boldsymbol {\bar X}}\|_2,
\end{align}
where ${\boldsymbol F}' =diag ({\boldsymbol F},{\boldsymbol F},{\boldsymbol F},{\boldsymbol F})\in {\mathbb C}^{4N_c \times 4N_c}$ is a block diagonal matrix in which the diagonal element is $\boldsymbol F$; $\boldsymbol {\bar X} = [{\boldsymbol{ \bar X}}_{ 0},{\boldsymbol {\bar X}}_{1},{\boldsymbol {\bar X}}_{2}, {\boldsymbol {\bar X}}_{3}] \in {\mathcal C}^{N_r \times 4N_c}$ is the pilot symbols. Accordingly, we have 
\begin{align}
\label{ESN_weights_MIMO}
\nonumber
{\boldsymbol W} &={\boldsymbol {\bar X}} [{\boldsymbol S}_{ 0}{\boldsymbol F},{\boldsymbol S}_{1}{\boldsymbol F},{\boldsymbol S}_{2}{\boldsymbol F},{\boldsymbol S}_{3}{\boldsymbol F}]^+ \\
&\stackrel {(a)}=  [{\boldsymbol {\bar X}}_{0} {\boldsymbol F}^H, {\boldsymbol {\bar X}}_{1}{\boldsymbol F}^H, {\boldsymbol {\bar X}}_{2}{\boldsymbol F}^H, {\boldsymbol {\bar X}}_{3}{\boldsymbol F}^H] 
[{\boldsymbol S}_{0},{\boldsymbol S}_{1}, {\boldsymbol S}_{2}, {\boldsymbol S}_{ 3} ]^+.
\end{align}
From $(a)$, we know that the weight learning can be conducted in the time domain as well.

Now, we consider the MIMO channel with a non-zero Doppler shift. To be compatible with the conventional pilots design in SISO depicted in the third sub-figure of Fig. \ref{fig_OFDM_pilots1}, we directly utilize this scattered pilots pattern as the training set of WESN. Therefore, the weights of the outputs are updated by
\begin{align}
\label{scatter_pilot_learning}
\min_{{\boldsymbol W}} \|{\boldsymbol W}[{\boldsymbol S}_{0}{\boldsymbol F}(:,{\Omega_{f_0}}), {\boldsymbol S}_{4}{\boldsymbol F}(:,{\Omega_{f_4}})] - [{\boldsymbol {\bar x}}^T_{0}(\Omega_{f_0}), {\boldsymbol {\bar x}}^T_{4}(\Omega_{f_4})]\|_2,
\end{align}
where $\Omega_{f_0}$ and $\Omega_{f_4}$ respectively represents the sub-carriers allocated to the pilot symbols at $t = 0$ and $t =4$ in the figure. Alternatively, the above minimization problem can be expressed as
\begin{align}
\label{scatter_pilot_learning2}
\min_{{\boldsymbol W}} \|{\boldsymbol W}[{\boldsymbol S}_{0}{\boldsymbol F}_{\Omega_{f_0}}, {\boldsymbol S}_{4}{\boldsymbol F}_{\Omega_{f_4}}] - [{\boldsymbol {\bar x}}^T_{0,{\Omega_{f_0}}}, {\boldsymbol {\bar x}}^T_{4,{\Omega_{f_4}}}]\|_2,
\end{align}
where 
\begin{align}
\label{def_sub_X}
{\boldsymbol {\bar x}}_{t,\Omega_{f}}(n)  \triangleq \begin{cases} {\boldsymbol {\bar x}}_{t}(n), & n \in \Omega_{f} \\0, & n \notin \Omega_{f} \end{cases} 
\end{align}

\begin{align}
\label{def_sub_F}
{\boldsymbol F}_{\Omega_{f}}(n)  \triangleq \begin{cases} {\boldsymbol F}(n), & n \in \Omega_{f} \\\boldsymbol 0, & n \notin \Omega_{f} \end{cases} .
\end{align}
Therefore, the output weight is given by
\begin{align}
{\boldsymbol W} = [{\boldsymbol {\bar x}}^T_{0,\Omega_{f_0}}, {\boldsymbol {\bar x}}^T_{4,\Omega_{f_4}}][{\boldsymbol S}_{0}{\boldsymbol F}_{\Omega_{f_0}}, {\boldsymbol S}_{4}{\boldsymbol F}_{\Omega_{f_4}}]^+,
\end{align}
which can be rewritten as
\begin{align}
{\boldsymbol W} = [{\boldsymbol {\bar x}}^T_{0,\Omega_{f_0}}{\boldsymbol F}^H, {\boldsymbol {\bar x}}^T_{4,\Omega_{f_4}}{\boldsymbol F}^H][{\boldsymbol S}_{0}{\boldsymbol F}_{\Omega_{f_0}}{\boldsymbol F}^H, {\boldsymbol S}_{4}{\boldsymbol F}_{\Omega_{f_4}}{\boldsymbol F}^H]^+,
\end{align}
where ${\boldsymbol {\bar x}}^T_{\Omega_{f}}{\boldsymbol F}^H$ represents the time domain OFDM waveform transformed merely from the symbols defined on the sub-carriers ${\bar \Omega_f}$. It demonstrates the output weight is also obtained by fitting the waveform of scattered pilots. Similarly, using the scattered pilots of MIMO illustrated in Fig. \ref{fig_ESN_MIMO_pilots2}, we have the following learning rule,
\begin{align}
{\boldsymbol W} = [{\boldsymbol {\bar X}}^T_{0,\Omega_{f_0}}{\boldsymbol F}^H, {\boldsymbol {\bar X}}^T_{4,\Omega_{f_4}}{\boldsymbol F}^H][{\boldsymbol S}_{0}{\boldsymbol F}_{\Omega_{f_0}}{\boldsymbol F}^H, {\boldsymbol S}_{4}{\boldsymbol F}_{\Omega_{f_4}}{\boldsymbol F}^H]^+,
\end{align}
where ${\boldsymbol {\bar X}}_{t, \Omega_f}$ represents the MIMO pilots which is similar to  (\ref{def_sub_X}).

\section{Complexity Analysis}
\label{complex_analysis}
In this section, we compare the computational complexity of the RC-based symbol detector to the conventional methods discussed in Sec. \ref{Conventional_Detection}, where the complexity is evaluated by floating-point operations per second (FLOPS). The discussion in this section is divided into SISO and MIMO. 
\subsection{SISO}
\subsubsection{Channel Estimation}
For the conventional methods of solving the channel estimation problem (\ref{CSI_estimation}), $g(\cdot)$ is assumed as a linear function. When $l(\cdot)$ is chosen as mean squared error (MSE), we branch the discussion according to the pilot patterns plotted in Fig. \ref{fig_OFDM_pilots} (a). For the comb pilots, the objective function in (\ref{CSI_estimation}) is rewritten as 
\begin{align}
\min_{\boldsymbol {\tilde h}} {\mathbb E}\|{\boldsymbol {\tilde y}}_{i} - {\bar {\boldsymbol x}}_{i}\odot{\boldsymbol {\tilde h}}\|_F^2,
\end{align}
where $\odot$ denotes the Hadamard product. From \cite{coleri2002channel}, we know that the solution is given by
\begin{align}
\label{CSI_estimation_conv}
{\boldsymbol {\tilde h}} = {\boldsymbol R}_{h y} {\boldsymbol R}_{yy}^{-1}{\boldsymbol y}_{i},
\end{align}
where ${\boldsymbol R}_{hy} = {\boldsymbol F}{\boldsymbol R}_{hh}{\boldsymbol F}^H{\bar{\boldsymbol X}_{i}}^H$, ${\boldsymbol X}_i = diag({\boldsymbol x}_i)$, ${\boldsymbol R}_{yy} = {\boldsymbol X}_{i}{\boldsymbol F}{\boldsymbol R}_{hh}{\boldsymbol F}^H{\boldsymbol X}^H_{i} + \sigma^2{\boldsymbol I}$, and ${\boldsymbol R}_{hh}$ is the channel covariance matrix. For the scattered pilots, the channel coefficients on the time-frequency grids allocated as pilots are calculated by
\begin{align}
\min_{{\boldsymbol h}({\Omega_f})} {\mathbb E}\|{\boldsymbol {\tilde y}}_{i}[{\Omega_f}] - {\bar {\boldsymbol X}}_{i}({\Omega_f}){\boldsymbol h}[{\Omega_f}]\|_F^2,
\end{align}
which has a closed-form solution as follows
\begin{align}
{\boldsymbol h}[{\Omega_f}] = {\boldsymbol R}_{h Y}({\Omega_f}) {\boldsymbol R}_{yy}({\Omega_f})^{-1}{\boldsymbol {\tilde y}}_{i}[{\Omega_f}],
\end{align}
where ${\boldsymbol R}_{hY}({\Omega_f}) = {\boldsymbol F}({\Omega_f},:){\boldsymbol R}_{hh}{\boldsymbol F}({\Omega_f},:)^H{\bar{\boldsymbol X}_{i}}({\Omega_f})^H$ and ${\boldsymbol R}_{yy}({\Omega_f}) = {\boldsymbol X}_{i}({\Omega_f}){\boldsymbol F}({\Omega_f},:){\boldsymbol R}_{hh}{\boldsymbol F}({\Omega_f},:)^H{\boldsymbol X}({\Omega_f},:)^H_{i} + \sigma^2{\boldsymbol I}$. Thereafter, the channels on the rest grids are inferred by the interpolation as discussed in \cite{dong2007linear}. Specifically, when the channel tap is assumed to be uncorrelated i.e., ${\boldsymbol R}_{hh} = {\boldsymbol I}$, we have 
\begin{align}
{\tilde h}[n] = {\bar x}_{i}^*[n]\cdot{\tilde y}_{i}[n]/(|{\bar x}_{i}[n]|^2+\sigma^2),
\end{align}	
where $n$ stands for the index of sub-carriers.

\subsubsection{Symbol Detection}
For the symbol detection problem (\ref{symbol_detection}), when $l(\cdot)$ is selected as MSE, we have
\begin{align}
\label{symbol_detection_standard}
\min_{{\boldsymbol x}_{i}} {\mathbb E}\|{\boldsymbol {\tilde y}}_{i} - { {\boldsymbol x}}_{i}\odot{\hat{{\boldsymbol h}}}_{i}\|_F^2.
\end{align}
When the transmission symbols are uncorrelated between sub-carriers, (\ref{symbol_detection_standard}) becomes
\begin{align*}
\min_{x_{i}[n]} \sum_{n = 0}^{N_c -1} {\mathbb E}|{\tilde y}_{i}[n] -x_{i}[n]{\hat h}_{i }[n] |^2,
\end{align*}
which has a following solution
\begin{align}
\label{symbol_detection2}
{\hat x}_{i}[n] =  {\hat{ h}}_{i}^*[n]*{\tilde y}_{i}[n]/(|{\hat{ h}}_{i}[n]|^2+\sigma^2).
\end{align}.

\subsubsection{Complexity}
For complexity analysis, we first review the FLOPS of standard matrix operations. Given two matrices ${\boldsymbol A}\in {\mathbb C}^{m \times n} $ and ${\boldsymbol B}\in {\mathbb C}^{n \times p} $, the matrix product ${\boldsymbol A}{\boldsymbol B}$ requires $N_{FLOPS}({\boldsymbol A}{\boldsymbol B}) = 2mnp$ for the summations and additions. For any invertible matrix $\boldsymbol C \in {\mathbb C}^{n \times n}$, FLOPS of the inverse is $N_{FLOPS}({\boldsymbol C}^{-1}) = n^3+n^2+n$. When $\boldsymbol C\in{\mathbb C}^{m\times n}$ is with full column rank, FLOPS of the MP-inverse ${\boldsymbol C}^+$ is given by $3mn^2 + 2n^3$. Therefore, for the comb pilot pattern in Fig. \ref{fig_OFDM_pilots2}, the FLOPS of the LMMSE channel estimation (\ref{CSI_estimation_conv}) is $2N_c^2$, in which the calculation of the covariance matrices ${\boldsymbol R}_{hy}$ and ${\boldsymbol R}_{yy}$ are ommited. In the symbol detection stage (\ref{symbol_detection2}), the FLOPS is proportional to $N_c$. Thus, the total FLOPS for the LMMSE channel estimation plus the symbol detection is on the scale of $\delta N_c^2+(1-\delta)N_c$, where $\delta$ represents the ratio of the pilot symbols to all the transmission symbols in the OFDM system. Moreover, when we consider the scattered pilot pattern in Fig. \ref{fig_OFDM_pilots3}, the complexity of interpolation needs to be included. For the standard linear interpolation method, the FLOPS is on the scale of $7N_c(1-\kappa)$, where $\kappa$ is the ratio of pilot sub-carriers over all sub-carriers. Thus, the total FLOPS for the LMMSE channel estimation with LMMSE symbol detection using scattered pilot is $\delta (\kappa N_c)^2+\delta7N_c(1-\kappa)+(1-\delta)N_c+\delta(1-\kappa) N_c$.

For the ESN/WESN using comb pilots, according to (\ref{ESN_output_weights}), the FLOPS for the output weights learning is $2N_c(N_n+1) + 3N_cN_n^2+2N_n^3$. Meanwhile, the computation at the symbol detection stage is merely on the output layer mapping, where the FLOPS is $N_nN_c$. Thus, the overall FLOPS for the ESN/WESN based symbol detection is $\delta(2N_c(N_n+1) + 3N_cN_n^2+2N_n^3) + (1-\delta) N_nN_c$. For scattered pilots, FLOPS at the learning stage is $2(\kappa N_c)(N_n+1)+ 3{\kappa N_c}N_n^2+2N_n^3$. Therefore, the total number of FLOPS is proportional to $\delta(2(\kappa N_c)(N_n+1)+ 3{\kappa N_c}N_n^2+2N_n^3) + N_nN_c $. It indicates the resulting complexity of the ESN/WESN receiver is linearly proportional to the number of subcarriers. It suggests that the ESN/WESN has less computational burden than the LMMSE method when the number of subcarriers is large. Remark that we do not consider the computations inside the reservoirs in this analysis. This is because the reservoirs are usually implemented through analog circuits which perform faster than the digital circuit\cite{duport2016fully,vandoorne2014experimental} with less energy consumption.

\subsection{MIMO}

By using the comb and scattered pilots respectively plotted in Fig. \ref{fig_OFDM_pilots2} and Fig. \ref{fig_OFDM_pilots3}, the FLOPS of the LMMSE channel estimation on each antenna pair is the same as the SISO case due to free interference. Therefore, the complexity of the MIMO channel estimation is $N_t N_r$ times to the SISO case. However, for the symbol detection, the interference caused by multiple transmitted antennas are required to be annihilated. Thus, the MIMO symbol detection demands more computations than the SISO case.

Now, we consider the LMMSE MIMO symbol detection using (\ref{symbol_detection}). When the transmitted symbols on different sub-carriers are independent to each others, the symbol detection can be conducted in sub-carrier-wise. Therefore, at the $n$th sub-carrier of the $t$th OFDM symbol, the symbol detection is solved by
\begin{align}
\min_{{\boldsymbol {\tilde x}}_{i}(n) } {\mathbb E}\|{\boldsymbol {\tilde y}}_{i}(n)- { \boldsymbol {\hat H}}_{i}(n){\boldsymbol {\tilde x}}_{i}(n)\|^2_F,
\end{align}
which has the following closed-form solution
\begin{align}
{\boldsymbol {\tilde x}}_{i}(n) = ({ \boldsymbol {\hat H}}_{i}^H(n){ \boldsymbol {\hat H}}_{i}(n)+\sigma^2\boldsymbol I)^{-1}{ \boldsymbol {\hat H}}_{i}^H(n){\boldsymbol {\tilde y}}_{i}(n).
\end{align}
It leads the FLOPS to $2N_c(N^3+N^2+N)$, where $N$ denotes the number of antennas at Tx and Rx when $N_t = N_r$. 

For the MIMO sphere decoding, it is an approximation of solving the following maximum likelihood estimation,
\begin{align}
\min_{{\boldsymbol x}_{i}(n) \in {\mathcal C}^{N_r}}\|{\boldsymbol {\tilde y}}_i(n) - {\boldsymbol {\hat H}}(n){\boldsymbol {\tilde x}}(n)\|_2,
\end{align}
where $\mathcal C$ represents the modulation constellation of the transmitted symbols. Since the standard sphere decoding usually has high redundancy in the implementation. We choose a complexity reduced sphere decoding algorithm proposed in \cite{chang2012algorithm} for the evaluation. It shows that the FLOPS is proportional to $N_c|\mathcal C|^N (2N^2+2N-1)$ which implies the sphere decoding is extremely complicated when a high order modulation is adopted. Using the comb pilot for ESN/WESN, the FLOPS for output weight learning is $2N^2N_c(N_n+1) + 3N_cNN_n^2+2N_n^3$ according to (\ref{ESN_weights_MIMO}), where the number of training OFDM symbols is the same as the transmission antennas. At the symbol detection stage, the FLOPS is $N_cNN_n$. Similarly, we can calculate the FLOPS using the scattered pilots. The results of complexity comparison is summarized in Table \ref{complexity_table}. We see that the computational complexity of ESN/WESN is dominated by the number of neurons which is smaller than $N_c$ through the numerical experiments in Sec. \ref{performance_evaluation}.
\begin{table}
	\centering
	\caption{Computational Complexity of Symbol Detection Methods}
	\label{complexity_table}
	\begin{tabular}{|c|c|}
		\hline
		Symbol Detection Method & Number of FLOPS\\
		\hline
		SISO LMMSE CSI with LMMSE  & $\delta (\kappa N_c)^2+\delta7N_c(1-\kappa)+(1-\delta)N_c+\delta(1-\kappa) N_c$\\
		\hline
		SISO ESN/WESN &$\delta(2(\kappa N_c)(N_n+1)+ 3{\kappa N_c}N_n^2+2N_n^3) + N_nN_c $\\
		\hline
		MIMO LMMSE CSI with LMMSE &
		$N^2(\delta (\kappa N_c)^2+\delta7N_c(1-\kappa))$ +
		$(1-\kappa\delta)N_c(N^3+N^2+N)$
		\\
		\hline
		MIMO LMMSE CSI with SD & $N^2(\delta (\kappa N_c)^2+\delta7N_c(1-\kappa)) + (1-\kappa \delta)N_c|\mathcal C|^N (2N^2+2N-1)$\\
		\hline
		MIMO ESN/WESN & $\delta(2N^2{\kappa N_c}(N_n+1) + 3\kappa N_cNN_n^2+2N_n^3)+N_cNN_n$ \\ 
		\hline
	\end{tabular}
	\par
	\bigskip
	$N$: the number of antennas at Tx and Rx; $\delta$: the ratio of OFDM pilots; $N_c$: the number of sub-carriers; $\kappa$: the ratio of pilots over all sub-carriers; $N_n$: the number of neurons and the length of buffer together. 
	\vspace{-8mm}
\end{table}

\section{Performance Evaluation}
\label{PE}
\label{performance_evaluation}
In this section, we evaluate the performance of the WESN based symbol detection. Through our numerical experiments, we incorporate the model of RF circuits, such as up/downsamplers, PA, and anti-interference/alias filters into the link simulation. To simulate the analog domain, we apply four times up-sampling upon the baseband signal. We assume that the channel is given by the following tap-delay model:
\begin{align}
\label{Channel_Model}
h_i^{(q, p)}(\tau) = \sum_{l = 0}^{L-1} a_i^{(p,q)}(l)p_w(\tau - \tau_l),
\end{align}
where $L$ is the maximum number of resolvable paths and $p_w(\tau)$ is the pulse shaping function which is chosen as the ideal rectangular shaped filter in the frequency domain. At the $l$th delay tap, we assume $a_i(l)$ is generated by the circular Gaussian distribution,
\begin{align}
\nonumber
a_i(l) &\sim {\mathcal {NC}}(0,\sigma_l^2 ),
\end{align}
where $\sigma_l^2$ is assumed to be an exponential power delay profile, i.e., $ \sigma_l^2 = exp(-\alpha\tau_l/\tau_{\max})$. Moreover, between two adjacent OFDM symbols, the correlation is assumed to be
\begin{align}
{\mathbb E}(a_i(l)a_{(i+1)}(l)) &= \sigma_l^2J_0(2\pi f_D \Delta t),
\end{align}
where $J_0$ stands for the Bessel function of the first kind with parameter $0$. Note that, for simplicity, we set the path-coefficients for any two different Tx-Rx antenna pairs to be independent. In general, other spatial correlation models or channel models also can be utilized without changing the training framework. The number of paths, $L$, in the channel model (\ref{Channel_Model}) is set as $6$.  The base-band modulation order is selected as 16-QAM. For the conventional methods using scattered pilots, the CSI is obtained by linear interpolation. 

Furthermore, it is important to note that the WESN/ESN symbol detector is trained using compatible pilot patterns of LTE/LTE-Advanced systems making it completely different from most of the existing literature, such as~\cite{He2018,Samuel2018,tan2018improving}. 
Almost all other work in the field assumes a large training set to train the underlying neural networks for symbol detection while we are focusing on using the extremely limited training overhead provided by LTE/LTE-Advanced systems. 
Since there are no obvious methods to extend these work to the limited training set we are investigating, the comparisons between our work and existing deep neural network-based approaches are not incorporated into our simulation evaluations. 
Rather, we compare our strategy with traditional signal processing based approaches using the same training overhead.

\subsection{Overfitting Issue}
\begin{figure}
	\centering
	\includegraphics[width=0.5\linewidth, height = 0.4\linewidth]{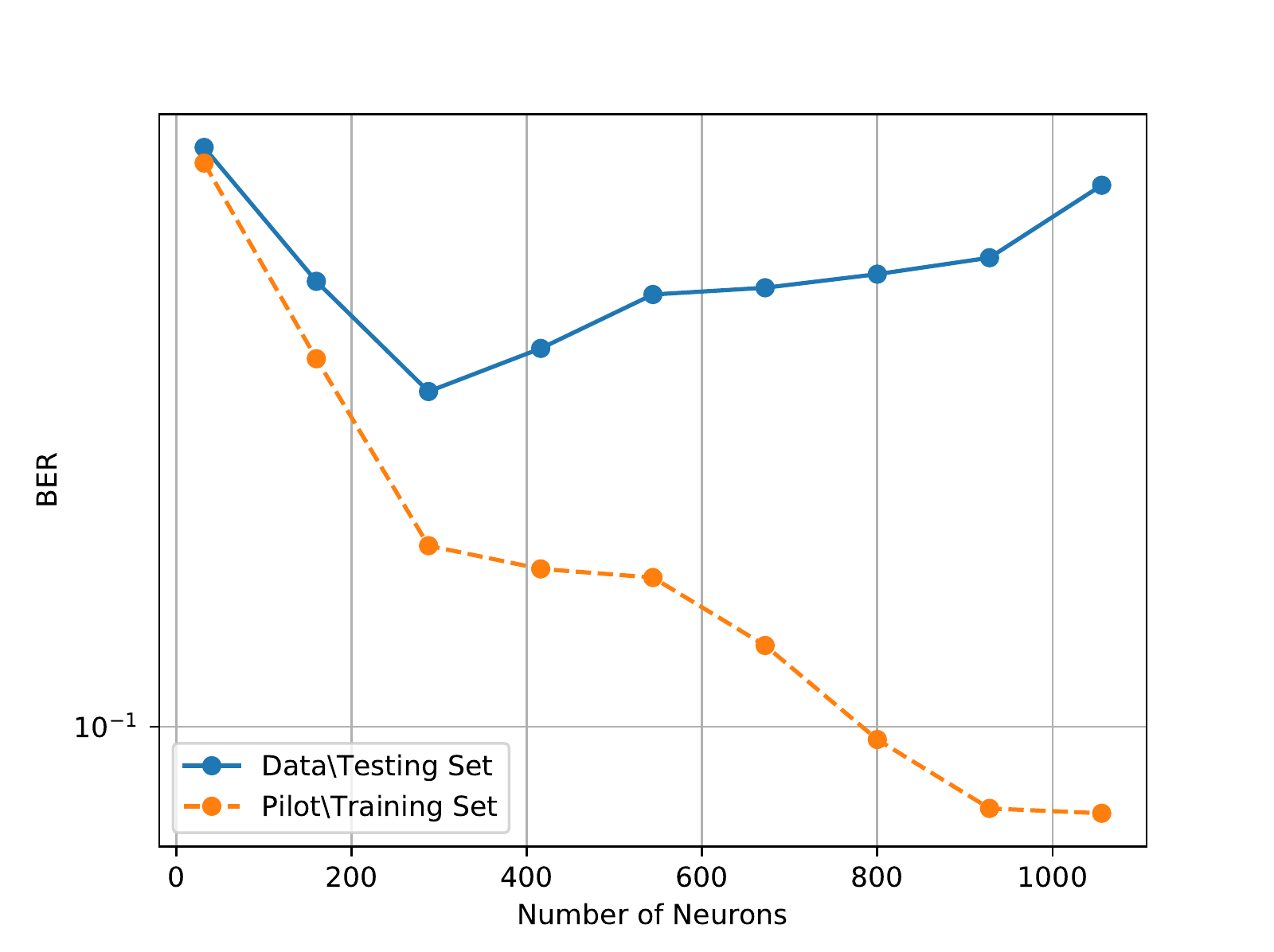}
	\vspace{-5mm}
	\caption{The over-fitting issue of changing the number of neurons in ESN under the MIMO block fading channel.}
	\label{overfitting}
	\vspace{-6mm}
\end{figure}

Before proceeding on the comparison between the RC based receiver and the conventional methods, we first reveal the overfitting issue on selecting the number of neurons/reservoirs of the underlying RC based receiver. As shown in Fig. \ref{overfitting}, we see that the BER of the training set decreases as the model becomes more complicated. At the same time, the BER gap between the testing set and training set is enlarged as the number of neurons increases. Therefore, in order to achieve low generalization error (i.e., low BER on testing set), it requires a proper selection on the number of neurons.

\subsection{SISO}
\begin{figure}[!t]
	\centering
	\includegraphics[width=0.5\linewidth, height = 0.4\linewidth]{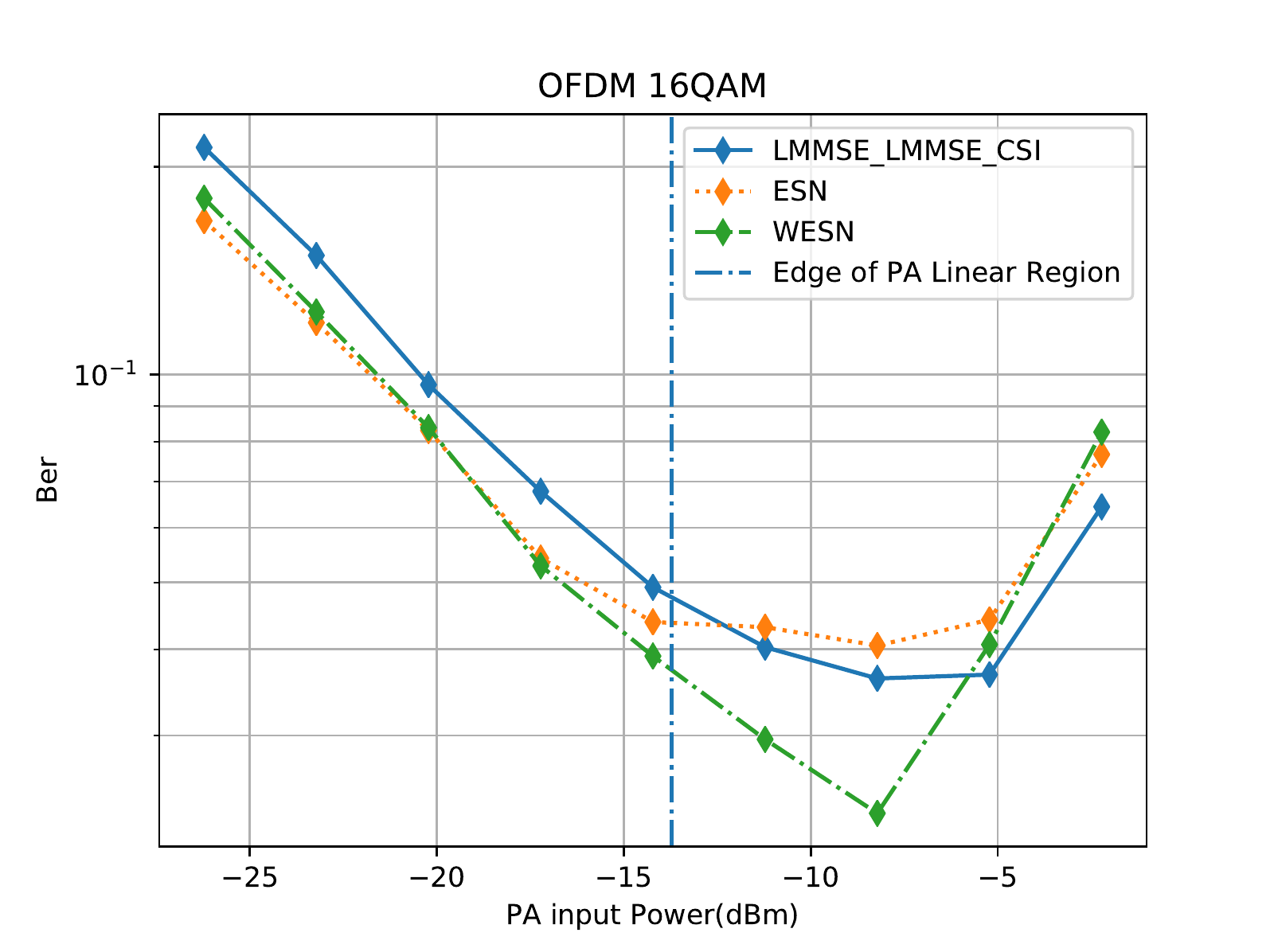}
	\vspace{-5mm}
	\caption{The BER comparison of the ESN symbol detector, WESN symbol detector and LMMSE method under the SISO block fading channel, where the number of neurons is set as 64 and the length of buffers is 30.}
	\label{FigSim1}
	\vspace{-6mm}
\end{figure}

We first evaluate the WESN receiver in the SISO channel under different operation regions of PA. Fig. \ref{FigSim1} shows the BER results when the Doppler shift is $0$Hz, where the threshold for PA linear region is set as 3dB up to the boundary of the linear region as depicted in Fig. \ref{PA_AM_AM}. Here the ESN is referred to as the buffer lenghth of WESN is set as $1$. The number of neurons for ESN and WESN is chosen to be the same, $64$. The buffer length of WESN is set as $30$. For, the labeled ``LMMSE-LMMSE-CSI'' method, the symbol detection is conducted by LMMSE using the CSI obtained from the LMMSE channel estimation. We can observe that these three methods have comparable performance among the linear region. Moreover, for WESN, the BER performance is the best in PA nonlinear region when the optimal PA input power is selected. It demonstrates that the WESN can considerably compensate for the non-linear waveform distortion. We can also conclude that the symbol detection using the estimated CSI does not necessarily lead to the optimality in BER performance. 

\begin{figure*}[!h]
	\centering 
	\subfloat[]{\includegraphics[width=0.49\linewidth, height = 0.4\linewidth]{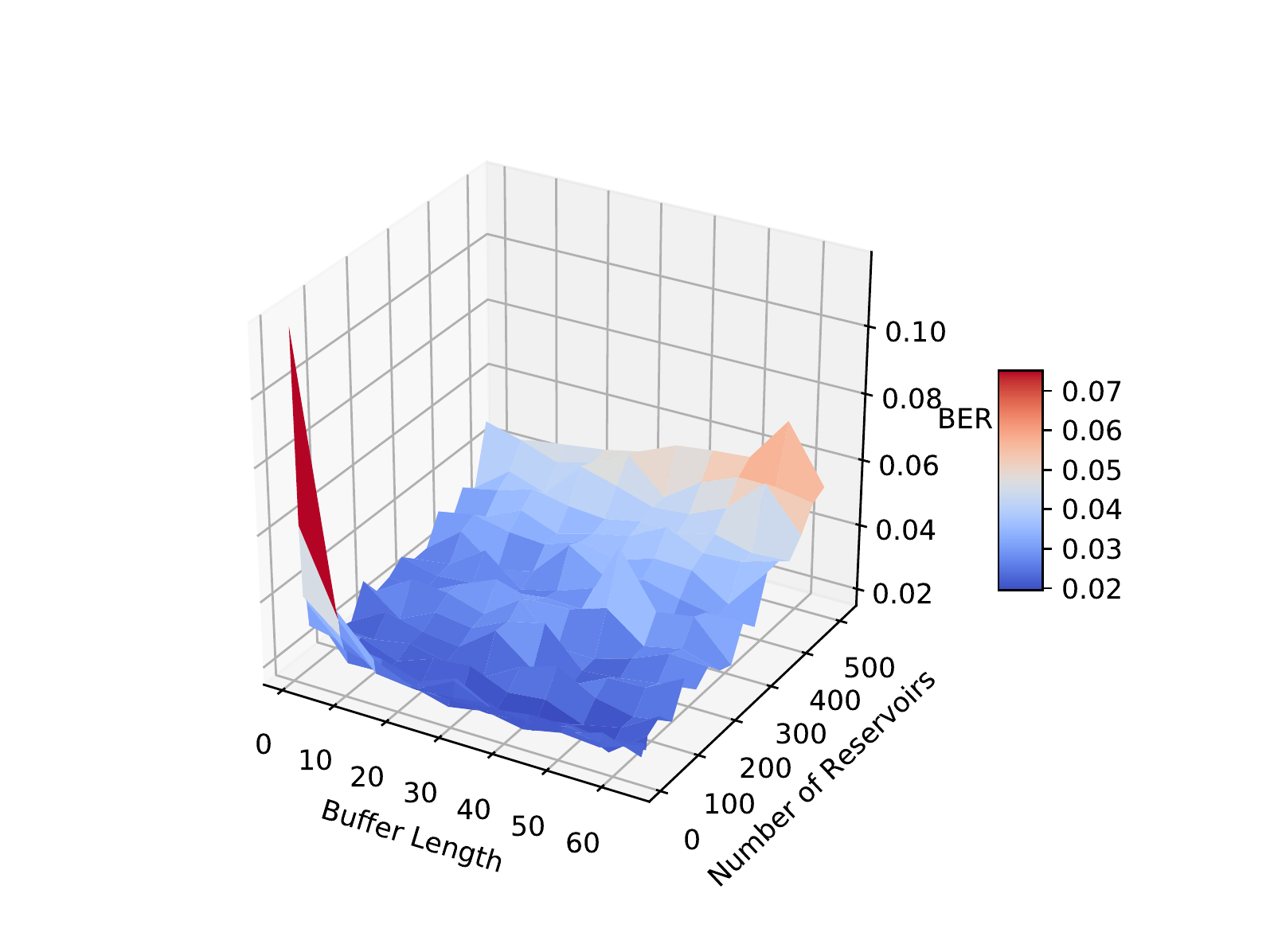}
		\label{fig_WESN_buffer_neuro}}
	\hfil
	\subfloat[]{\includegraphics[width=0.49\linewidth, height = 0.4\linewidth]{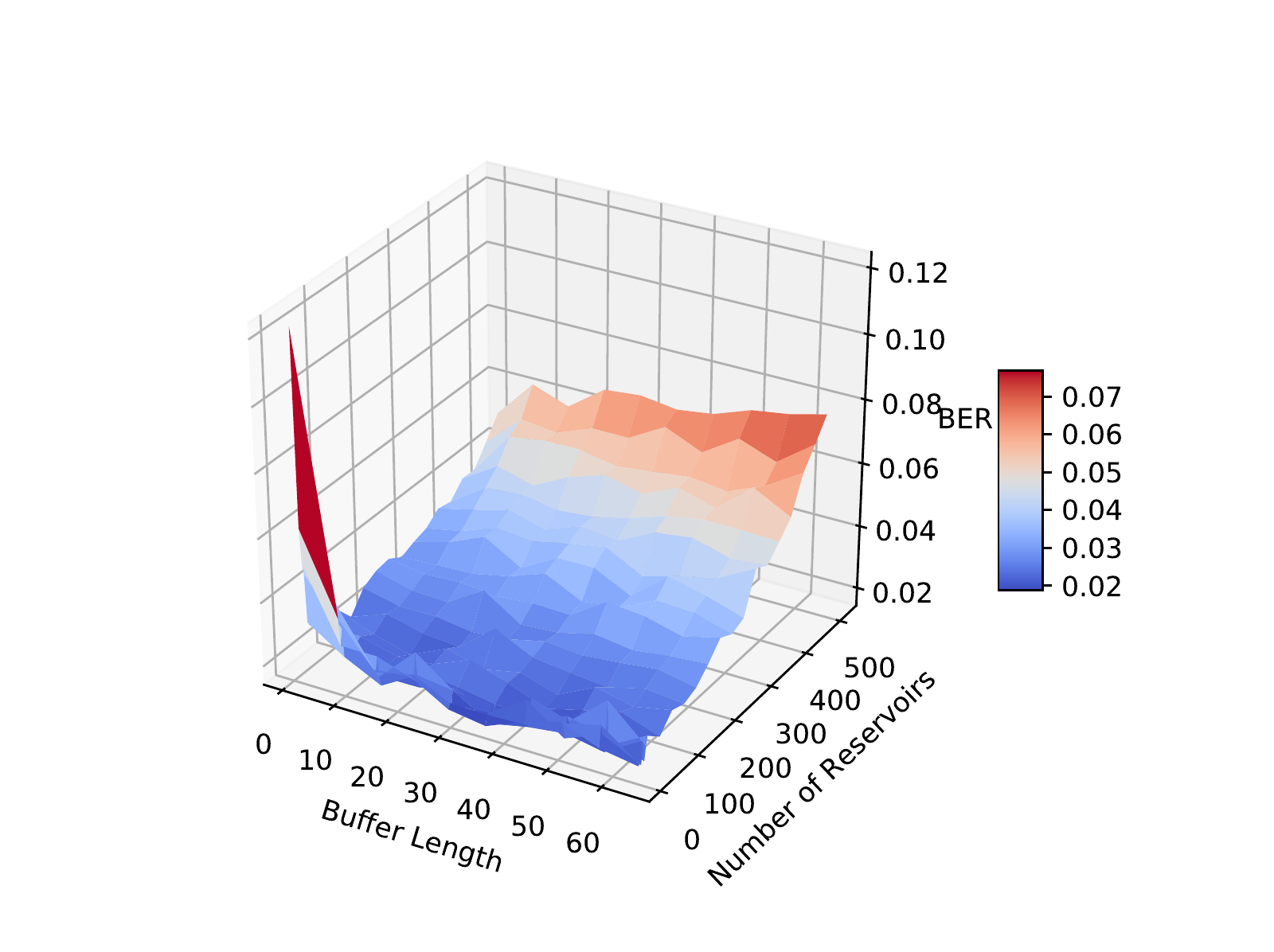}
		\label{fig_WESN_buffer_neuro1}}
	\hfil
	\caption{The average BER performance of the WESN symbol detector under the SISO block fading channel by varying the length of buffers and the number of neurons:(a) 3D surface when the PA input power is -8 dBm, (b) 3D surface when the PA input power is -11 dBm, where the number of neurons varies from 8 to 512 and the length of buffers ranges from 1 to 64.} 
	\label{fig_WESN_buffer_neuro_}
	\vspace{-5mm}
\end{figure*}

Nevertheless, the performance of the WESN receiver is highly related to the settings of neural network parameters, especially the number of internal reservoirs and the buffer length. We further investigate how the length of buffer and the number of neurons can jointly impact the BER performance. In Fig. \ref{fig_WESN_buffer_neuro_}. we observe that the length of buffer brings another degree of freedom to improve the symbol detection performance. From this figure, it shows that by either increasing the number of neurons or the length of buffer, the resulting BER declines. However, due to overfitting, BER increases again when the number of neurons becomes greater. Furthermore, it shows that compared to the WESN configured with more neurons, the WESN with a few numbers of neurons but longer buffers can achieve the same performance. This is because the memory capacity of WESN is jointly determined by the configuration of neurons and buffers. 
Furthermore, we see the overfitting issue in Fig. \ref{fig_WESN_buffer_neuro1} is slightly different from that in Fig. \ref{fig_WESN_buffer_neuro}. 
When the number of neurons is large enough (such as close to 500), the BER in Fig.~\ref{fig_WESN_buffer_neuro1} is higher than that in Fig. \ref{fig_WESN_buffer_neuro}. 
This is because when the input power is closer to the linear region, the transmitted signal is less distorted. 
Therefore, the size of the employed neural network is expected to be smaller. 
On the other hand, using more neuron states (more complicated models) can result in worse BER performance (overfitting) when the input power is close to the linear region.

\begin{figure}
	\centering
	\includegraphics[width=0.5\linewidth, height = 0.4\linewidth]{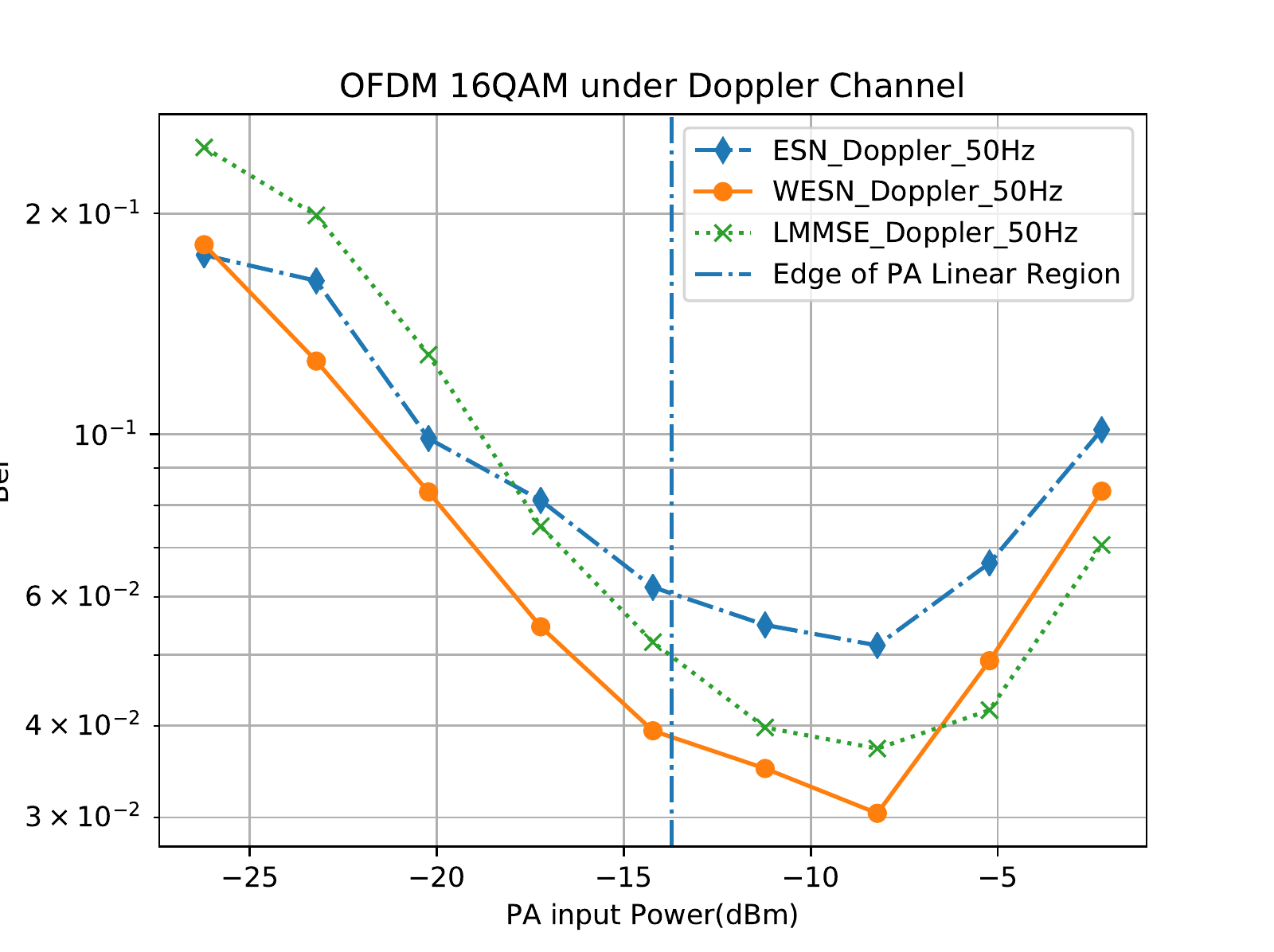}
	\vspace{-5mm}
	\caption{The BER comparison of the ESN symbol detector, the WESN detector and the LMMSE method under the SISO Doppler channel with different Doppler shifts, where the number of neurons is set as $64$ and the length of buffers is $30$.}
	\label{FigSim2}
	\vspace{-8mm}
\end{figure}

The BER performance under different Doppler shifts in the SISO channel is shown in Fig. \ref{FigSim2}. We see that these three methods are comparable in the BER as well. 

\begin{figure}
	\centering
	\includegraphics[width=0.5\linewidth, height = 0.4\linewidth]{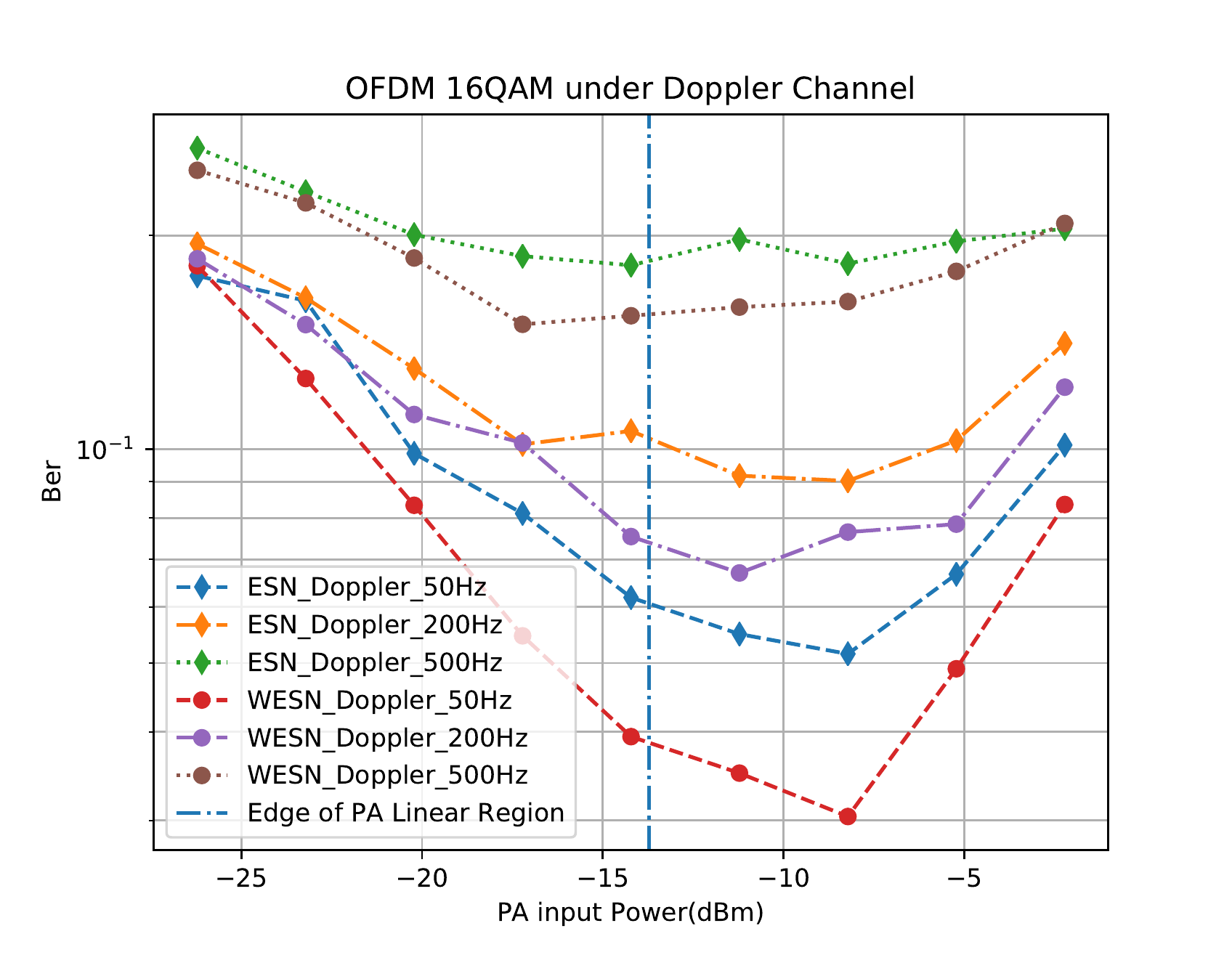}
	\vspace{-5mm}
	\caption{The BER comparison between the ESN symbol detector and the WESN symbol detector under SISO Doppler channel with different Doppler shifts, where the number of neurons is set as $64$ and the length of buffers is $30$.}
	\label{FigSim3}
	\vspace{-5mm}
\end{figure}

\begin{figure*}[!h]
	\centering 
	\subfloat[]{\includegraphics[width=0.49\linewidth, height = 0.4\linewidth]{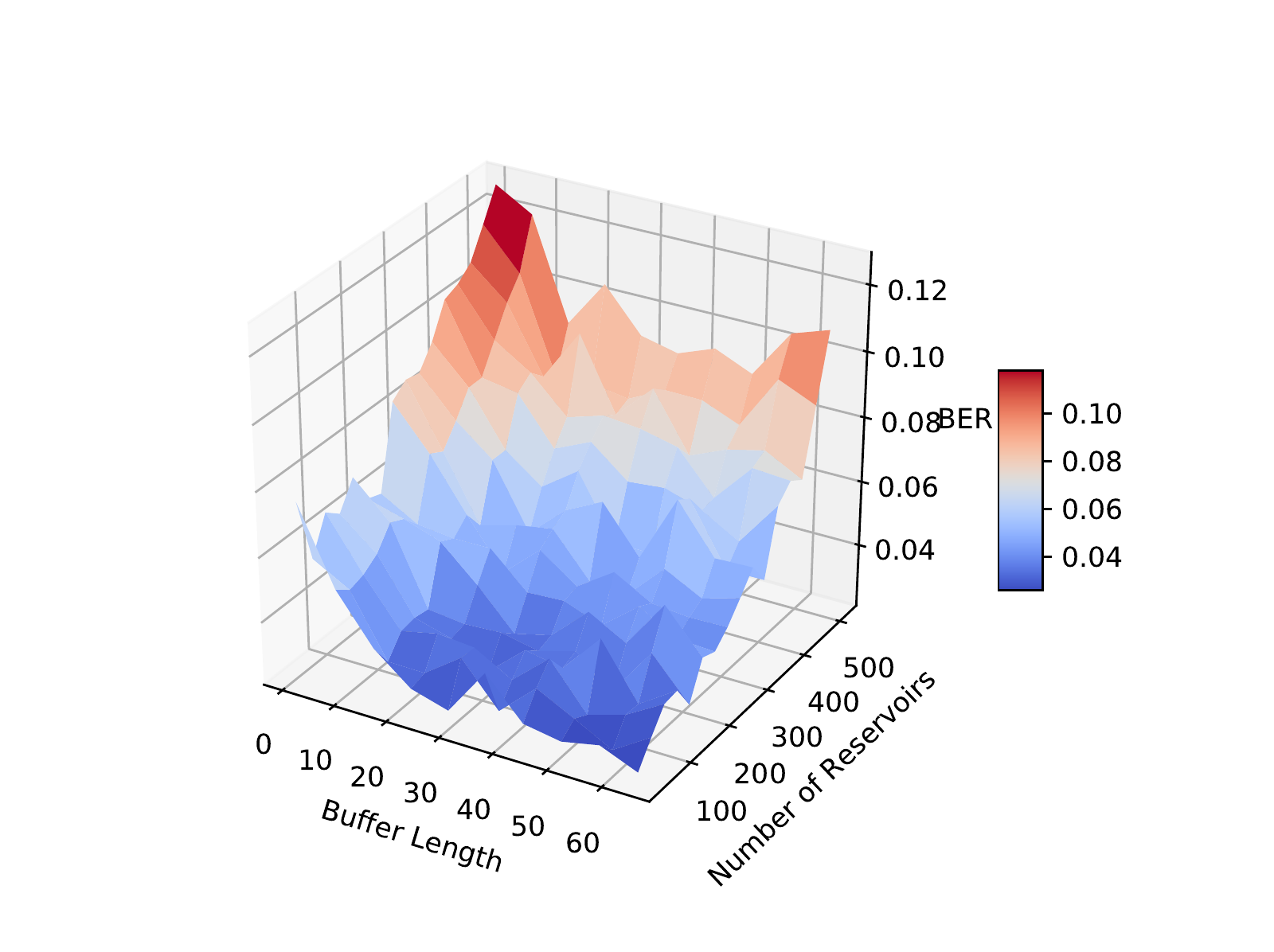}
		\label{fig_WESN_buffer_neuro_Doppler}}
	\subfloat[]{\includegraphics[width=0.49\linewidth, height = 0.4\linewidth]{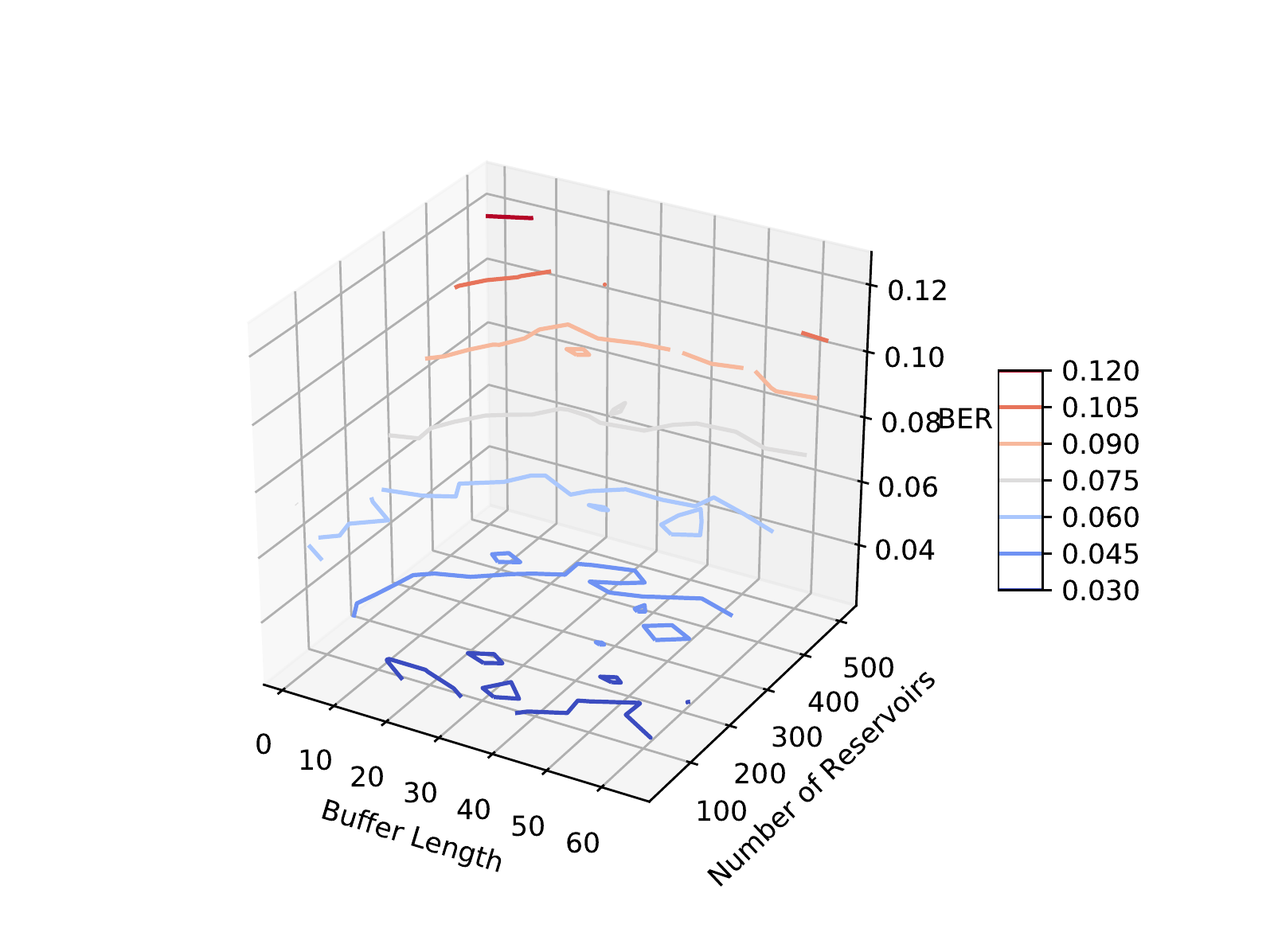}
		\label{fig_WESN_buffer_neuro_Doppler2}}
	\caption{The average BER performance of the WESN symbol detector under the SISO Doppler channel by varying the length of buffers and the number of neurons when the PA input power is -8 dBm: (a) 3D surface (b) 3D contour version, where the number of neurons varies from 8 to 512, the length of buffers ranges from 1 to 64 and the Doppler shift is 50 Hz.} 
	\label{fig_WESN_buffer_neuro_Doppler_}
	\vspace{-8mm}
\end{figure*}

In Fig. \ref{FigSim3}, the comparison between ESN and WESN under different Doppler shifts is investigated. We can observe that the WESN always perform better than the ESN under different Doppler shifts. From Fig. \ref{fig_WESN_buffer_neuro_Doppler_}, we again investigate the BER distribution by varying buffer length and neurons number. We see that increasing buffer size can significantly decrease BER which indicates that WESN can gain more advantages over the Doppler shift channel compared to the standard ESN. Meanwhile, adding more neurons can always lead to model overfitting.

\subsection{MIMO}

\begin{figure}
	\centering
	\includegraphics[width=0.5\linewidth, height = 0.4\linewidth]{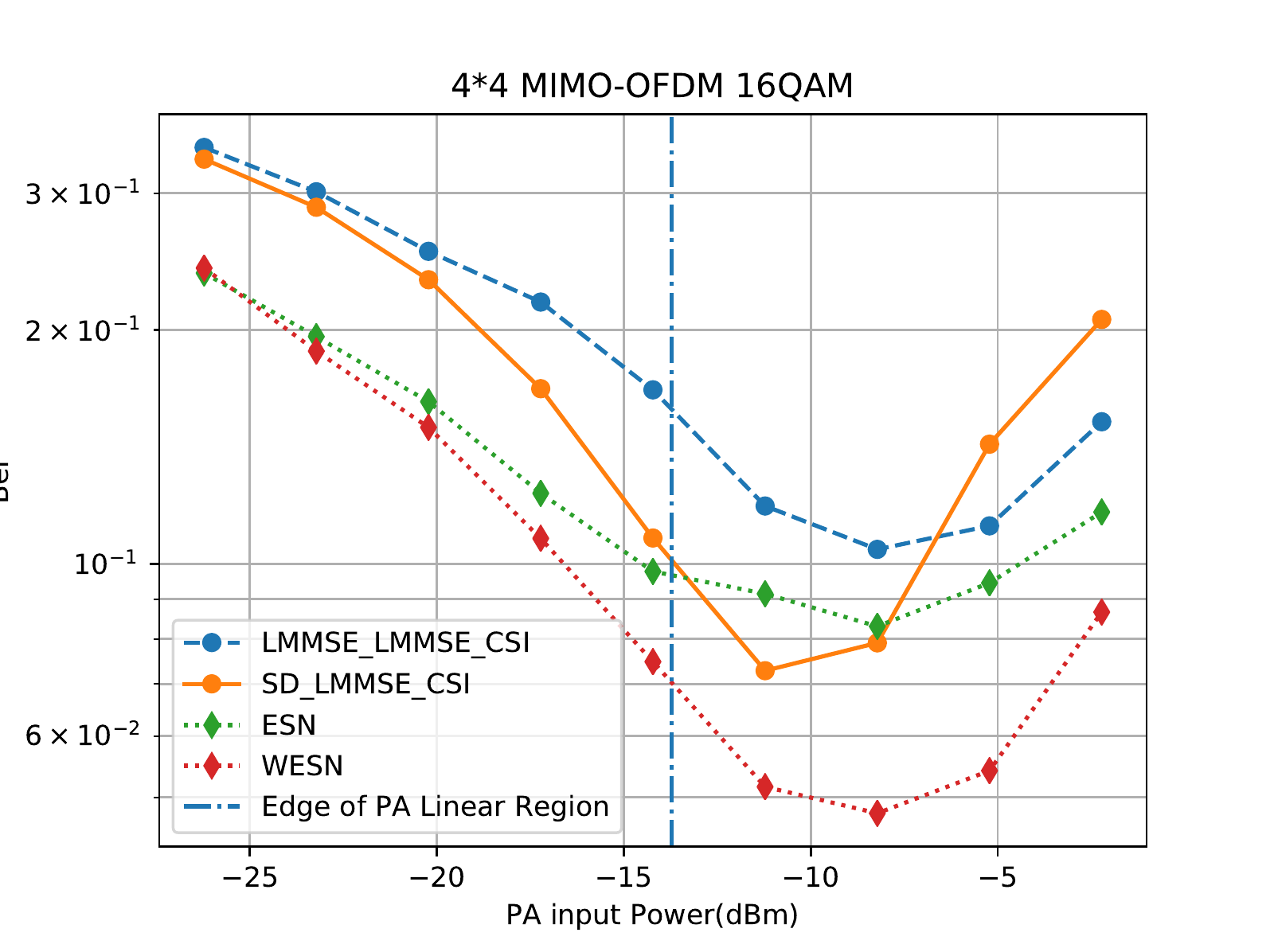}
	\vspace{-5mm}
	\caption{The BER comparison of the ESN symbol detector, the WESN symbol detector, the LMMSE method and sphere decoding under the MIMO block fading channel, where the number of neurons is set as 64 and the length of buffers is 30.}
	\vspace{-5 mm}
	\label{FigSim5}
\end{figure}

In Fig. \ref{FigSim5}, we compare the BER performance of WESN to the conventional methods, i.e., LMMSE and sphere decoding (SD) under block fading channel. For the conventional methods, the CSI is obtained by LMMSE using the pilot patterns depicted in Fig. \ref{fig_OFDM_pilots3}. We see that the performance gap between WESN and the conventional methods is enlarged compared to the SISO case. Especially, for SD, the BER performance deteriorates quickly when the PA input power is in the non-linear region. This is because SD requires more accurate CSI for symbol detection. 

\begin{figure}
	\centering 
	\subfloat[]{\includegraphics[width=0.49\linewidth, height = 0.4\linewidth]{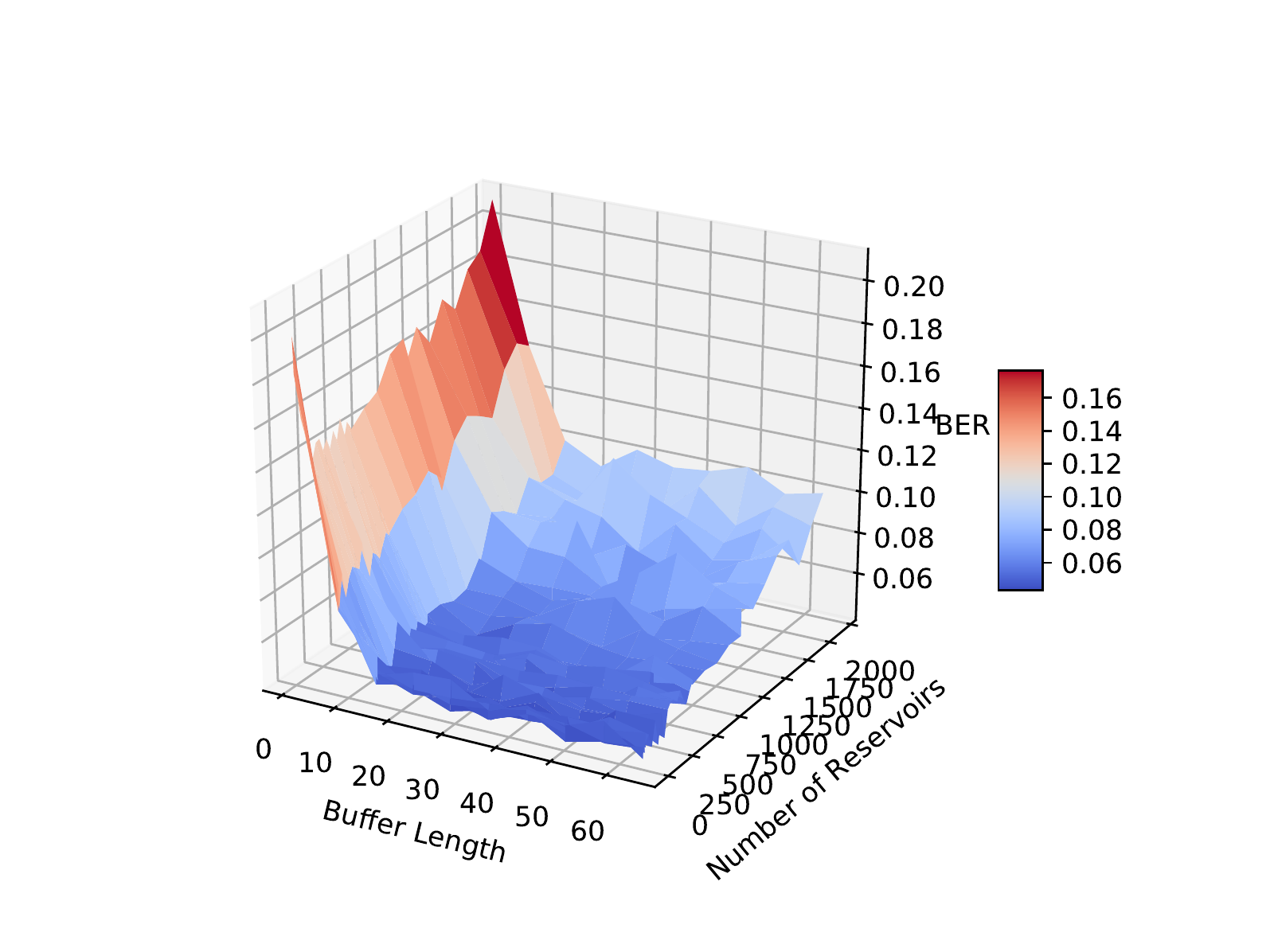}
		\label{fig_WESN_buffer_neuro_MIMO}}
	\hfil
	\subfloat[]{\includegraphics[width=0.49\linewidth, height = 0.4\linewidth]{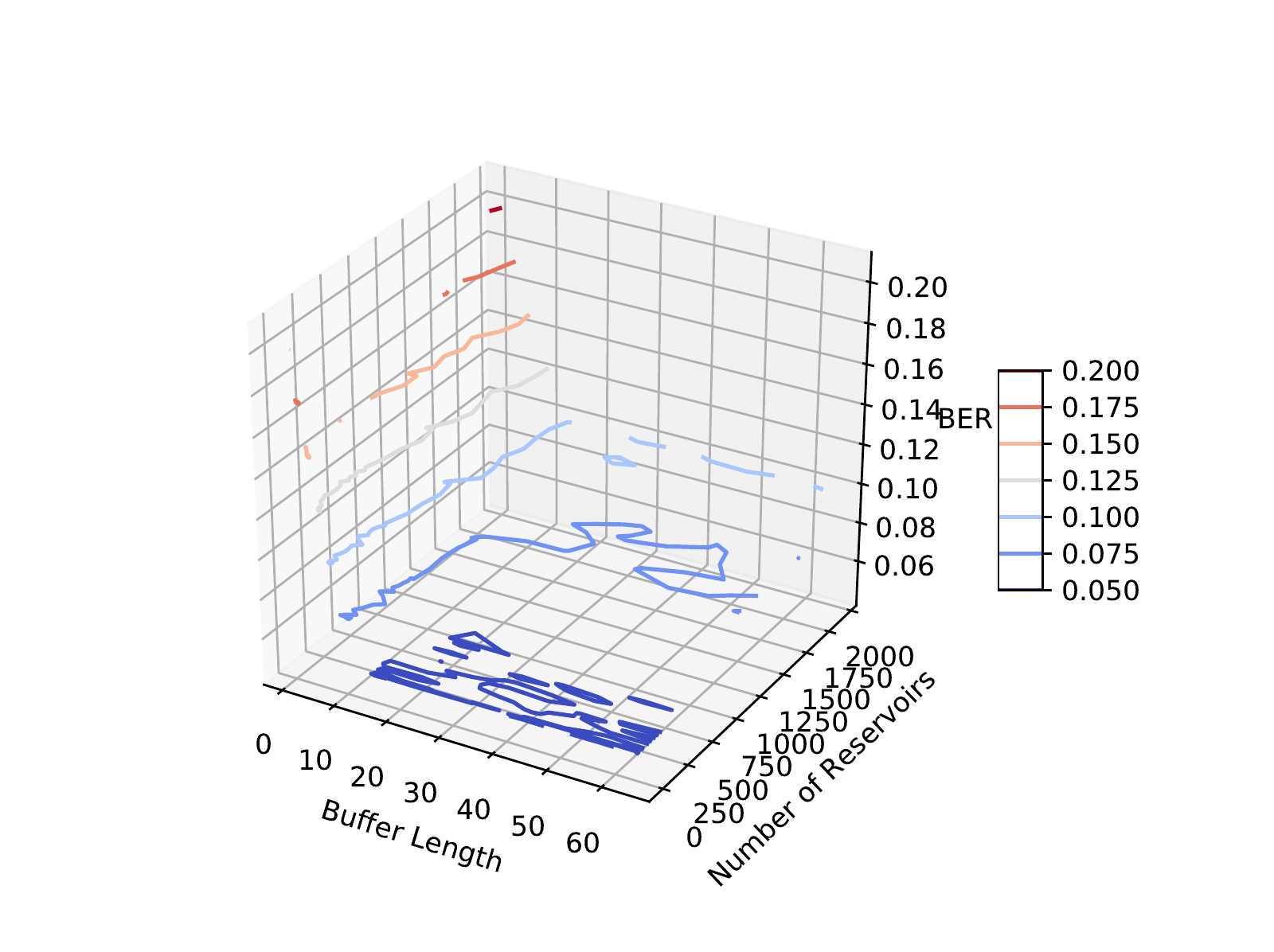}
		\label{fig_WESN_buffer_neuro_MIMO1}}
	\caption{The average BER performance of the WESN symbol detector under the MIMO block fading channel by varying the length of buffers and the number of neurons when the PA input power is -8 dBm: (a) 3D surface (b) 3D contour version, where the number of neurons varies from 8 to 512 and the length of buffers ranges from 1 to 64.}
	\vspace{-2 mm}
	\label{fig_WESN_buffer_neuro_MIMO_}
\end{figure}

\begin{figure}
	\centering 
	\includegraphics[width=0.5\linewidth, height = 0.4\linewidth]{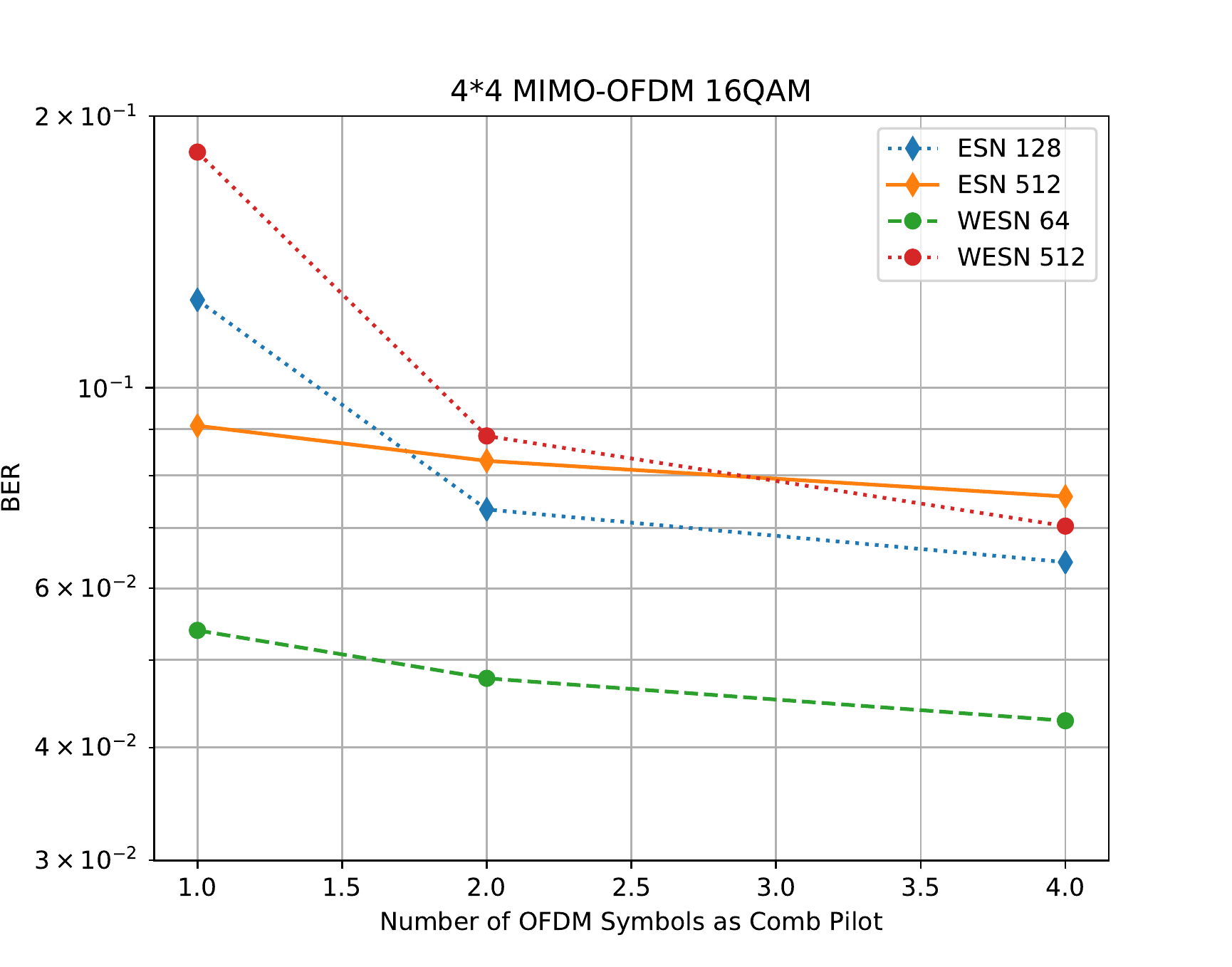}
	\label{fig_WESN_MIMO_block2}
	\vspace{-5mm}
	\caption{The BER performance of the ESN symbol detector and the WESN symbol detector under the MIMO block fading channel by varying the number of pilots OFDM symbols, where the PA input power is chosen as -9dBm, the number of neurons for ESN is equal to 128 and 512, the number of neurons for WESN is equal to 64 and 512 and the length of buffers is 30.}
	\label{pilot_lengths_BER}
	\vspace{-5mm}
\end{figure}

\begin{figure}
	\centering 
	\includegraphics[width=0.5\linewidth, height = 0.4\linewidth]{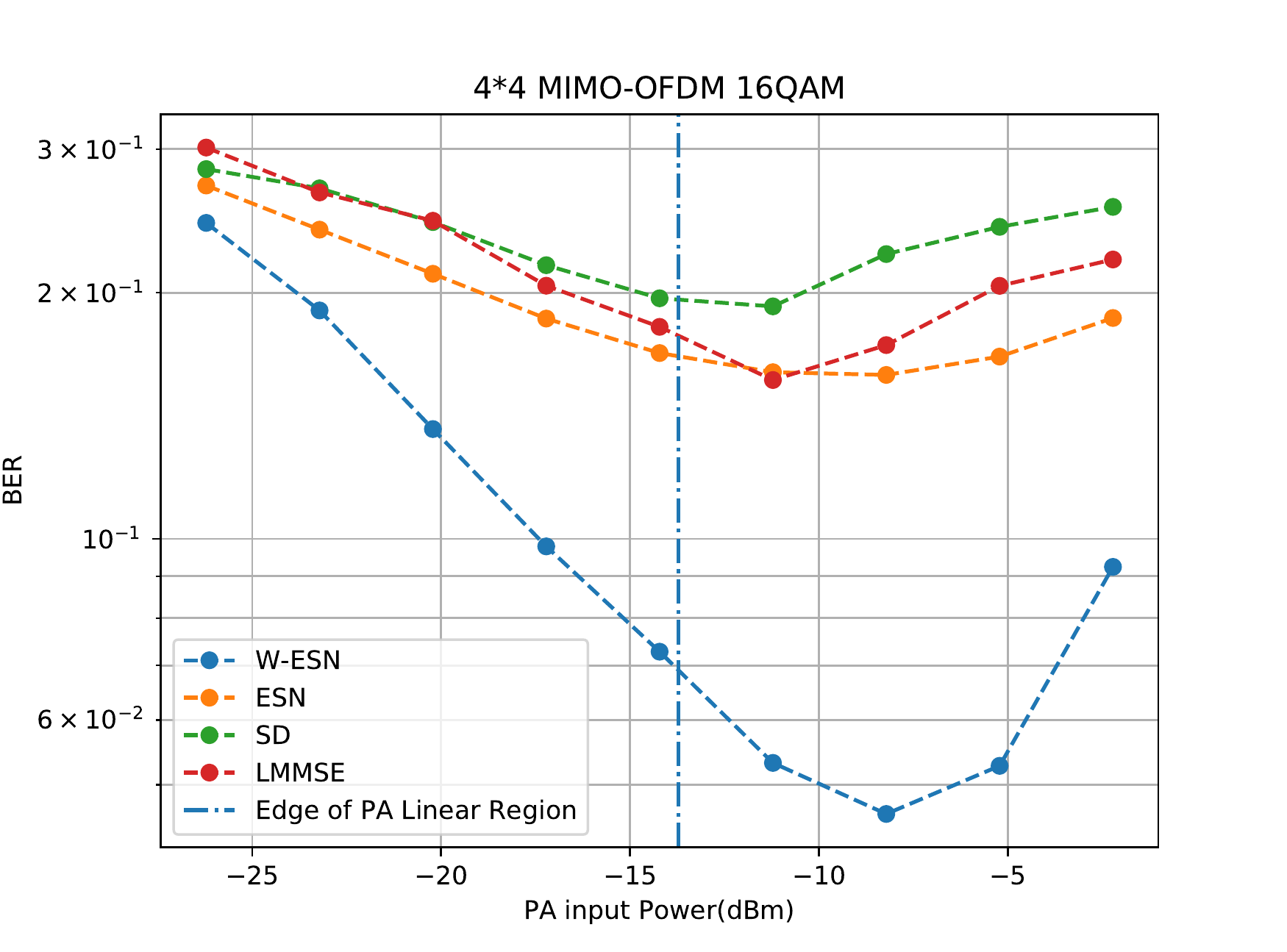}
	\vspace{-5mm}
	\caption{The BER comparison of the ESN symbol detector, the WESN symbol detector, the LMMSE method and sphere decoding under the MIMO Doppler channel, where the length of buffers is $30$, the number of neurons is $64$ and the Doppler shift is $50$Hz.}
	\vspace{-6 mm}
	\label{fig_WESN_MIMO_doppler}
\end{figure}

\begin{figure}
	\centering 
	\subfloat[]{\includegraphics[width=0.49\linewidth, height = 0.4\linewidth]{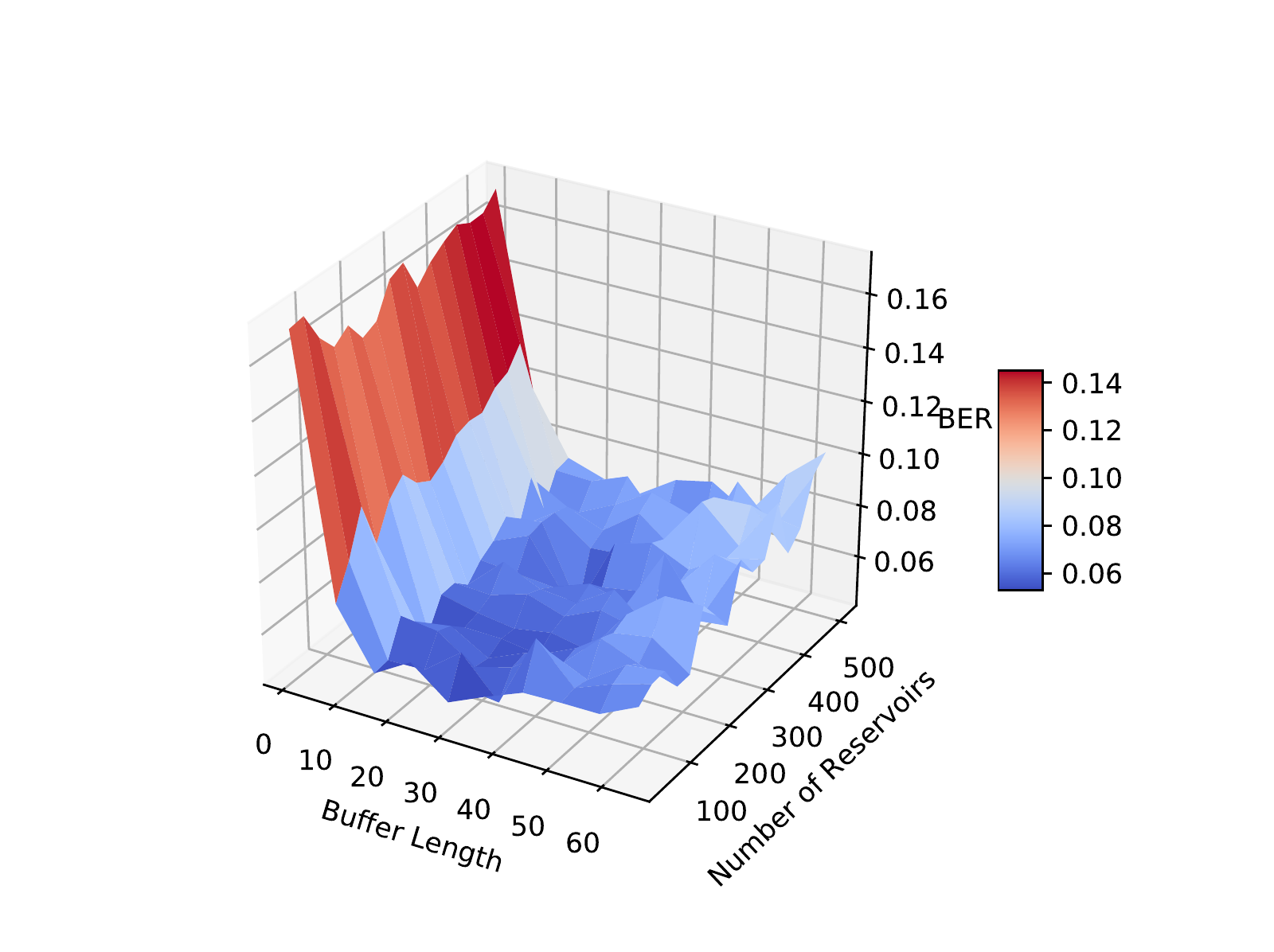}
		\label{fig_WESN_buffer_neuro_MIMO_doppler}}
	\hfil
	\subfloat[]{\includegraphics[width=0.49\linewidth, height = 0.4\linewidth]{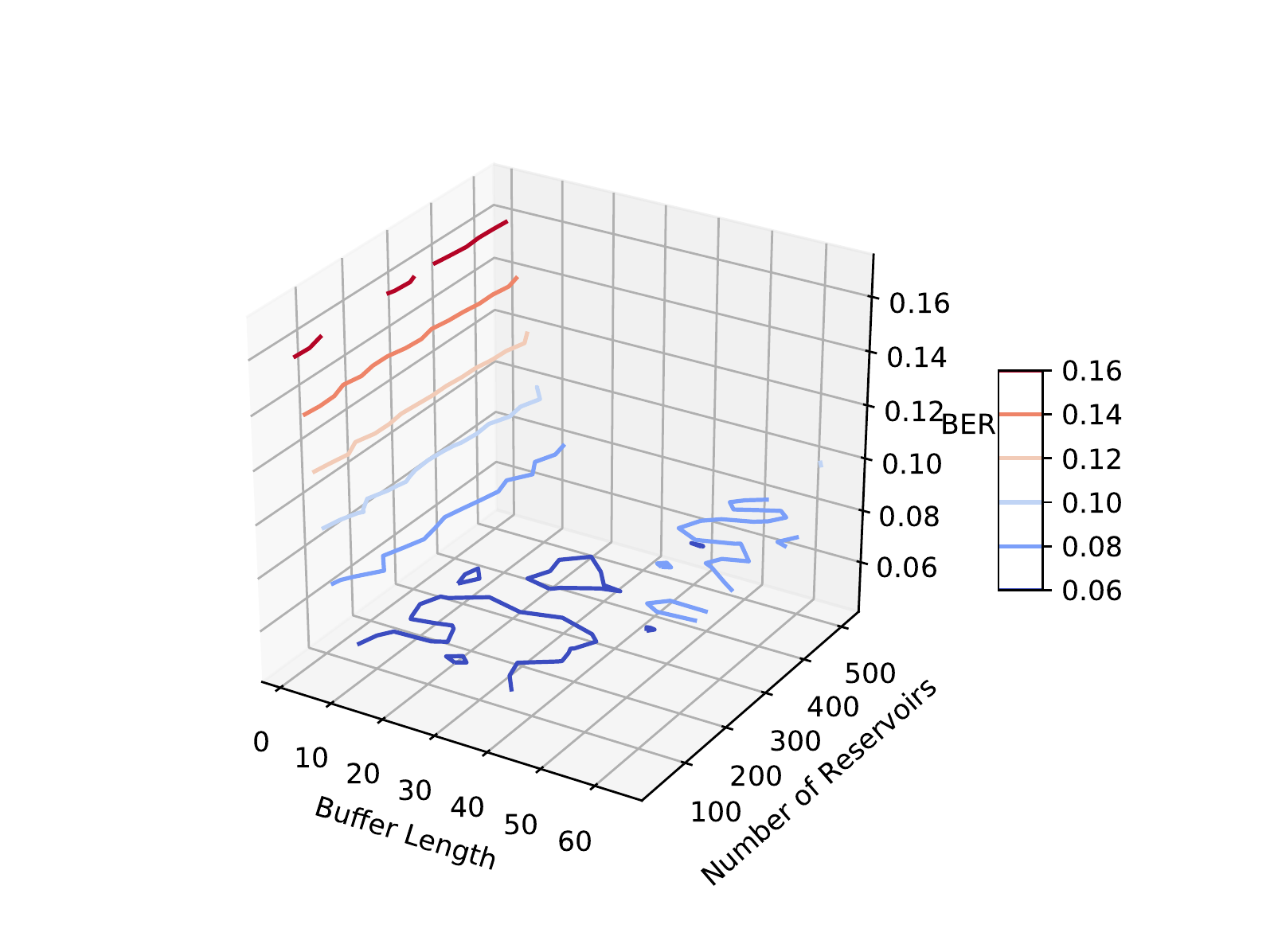}
		\label{fig_WESN_buffer_neuro_MIMO_doppler2}}
	\vspace{-2 mm}
	\caption{The average BER performance of the WESN symbol detector under the MIMO Doppler channel by varying the length of buffers and the number of neurons when the PA input power is $-8$ dBm: (a) 3D surface, (b) 3D contour version, where the number of neurons varies from $8$ to $512$ and the length of buffers ranges from $1$ to $64$ and the Doppler shift is $50$ Hz.}
	\vspace{-6 mm}
	\label{fig_WESN_buffer_neuro_MIMO_doppler_}
\end{figure}

Again, we plot the BER distribution by varying the buffer length and the number of neurons as shown in Fig. \ref{fig_WESN_buffer_neuro_MIMO_}. The advantages of the introduced buffer are more obvious compared to the SISO case by looking at Fig. \ref{fig_WESN_buffer_neuro_}. 
Moreover, by using the pilot pattern in Fig.\ref{fig_ESN_MIMO_pilots1}, the number of pilot symbols in training can be flexibly adjusted. In Fig. \ref{pilot_lengths_BER}, we show the BER performance by varying the number of pilots, i.e., the number of OFDM symbols allocated as pilots. To be clarified, Fig. \ref{fig_ESN_MIMO_pilots1} shows the number of OFDM pilot symbols is equal to $4$. Specifically, when $T<4$, it is non-orthogonal pilots as the number of pilot OFDM symbols is smaller than the number of Tx antennas, $4$. When we employ the conventional methods, using non-orthogonal pilot is not enough to avoid the pilot interference during the channel estimation stage. This means the conventional channel estimation method cannot be directly applied using non-orthogonal pilots. However, by using RC based method, we can observe that the BER performance is almost invariant compared to orthogonal pilots. It is because that the learning-based symbol detection can extract important features underlying the channel which are the inherent sparsity in the time-delay domain. Meanwhile, by increasing the number of neurons, we can also observe the deterioration of the BER performance due to overfitting.
In Fig. \ref{fig_WESN_MIMO_doppler}, we plotted the performance using the scattered pilot of MIMO under the Doppler shift channel. The 2D BER distribution under the Doppler channel is shown in Fig. \ref{fig_WESN_buffer_neuro_MIMO_doppler_} which has a similar distribution as Fig.  \ref{fig_WESN_buffer_neuro_MIMO_}.

Thus, we conclude that the RC based approach performs better than conventional methods in low SNR regime and nonlinear distortion channel. Furthermore, we want to highlight the training resources used by the introduced approach. As shown in the 3rd sub-figure of Fig. 2(a), the training set (demodulation reference signals) for SISO-OFDM is around $5\%$ of all the REs: one resource block has $12\times7=84$ REs with 4 of them being reference signals. On the other hand, for $4\times4$ MIMO-OFDM systems, one resource block has $12\times7\times4=336$ REs with 16 of them being reference signals (the overhead is around $20\%$). From the best of our knowledge, these training sets are too small to train a proper-fitted deep neural network from scratch. However, experiments show that the RC-based approach can achieve a good generalization result using such a few numbers of training.

\section{Conclusion}
\label{conclusion}
In this paper, we considered the application of reservoir computing, a special RNN, to MIMO-OFDM symbol detection. Compared to our previous work \cite{mosleh2017brain}, a new RC based detector, WESN, is introduced as the receiver to significantly improve the performance of interference cancellation. As an improvement upon previous ESN, the WESN is proved to be able to fundamentally enhance the short term memory of the reservoir computing system. Additionally, compared to conventional coherent MIMO-OFDM symbol detection strategies as well as ESN, numerical evaluation demonstrates that WESN offers great performance improvement even under the constraint of using compatible pilot patterns defined in 3GPP LTE standards in both static and dynamic MIMO channel. Moreover, through complexity analysis, we prove that WESN performs relatively few FLOPS compared with conventional methods. 
For future work, this symbol detection framework can be extended to learn soft demodulation information which can be utilized for joint symbol demodulation and channel decoding. 
Furthermore, it will be interesting to explore other activation functions with more complicated neural network architecture such as extending the shallow RC architecture to deep RNNs to further improve the detection performance.
 
\section*{Appendix}
\label{Meomery_ESN}
\subsubsection*{Proof of Theorem \ref{theorem2}}

We know the output weights for the $m$-th delay capacity can be calculated by 
\begin{align}
\min_{{\boldsymbol  w}_{out}} \|\tilde{\boldsymbol { x}}(m:N-1) - \tilde{\boldsymbol y}(0:N-m-1)\|_2^2.
\end{align}
Suppose ${\tilde {\boldsymbol x}} = {\boldsymbol w}_{out}{\boldsymbol S}_{WESN}$, where ${\boldsymbol S}_{WESN} = [[\tilde{\boldsymbol y}^T(m:m-M), {\tilde{\boldsymbol s}}^T(m)]^T, [\tilde{\boldsymbol y}^T(m+1:m+1-M), {\tilde{\boldsymbol s}}^T(m + 1)],\cdots,[\tilde{\boldsymbol y}^T(N - 1-M:N), {\tilde {\boldsymbol s}}^T(N - 1)]]^T$ represents the extended states as introduced in \cite{jaeger2001echo}. By splitting ${\boldsymbol w}_{out}$ into $[{\boldsymbol w}_1,{\boldsymbol w}_2]$, we have
\begin{align}
&\|[{\boldsymbol w}_1,{\boldsymbol w}_2][{\boldsymbol { Y}}^T, {\boldsymbol { S}}_{ESN}^T]^T - \tilde{\boldsymbol y}(0:N-m-1)\|_2^2  \\
&= \|{\boldsymbol w}_1{\boldsymbol { Y}} - \lambda\tilde{\boldsymbol y}(0:N-m-1)+{\boldsymbol w}_2{\boldsymbol { S}}_{ESN} - (1-\lambda)\tilde{\boldsymbol y}(0:N-m-1)\|_2^2\\
&\leq 2\|{\boldsymbol w}_1{\boldsymbol { Y}} - \lambda\tilde{\boldsymbol y}(0:N-m-1)\|_2^2+2\|{\boldsymbol w}_2{\boldsymbol { S}}_{ESN} - (1-\lambda)\tilde{\boldsymbol y}(0:N-m-1)\|_2^2,
\end{align}
where $\lambda \in (0, 1)$. Thus,
\begin{align}
&\min_{{{\boldsymbol w}_1,{\boldsymbol w}_2}}{1\over 2} \|[{\boldsymbol w}_1,{\boldsymbol w}_2][{\boldsymbol { Y}}^T, {\boldsymbol { S}}_{ESN}^T]^T - \tilde{\boldsymbol y}(0:N-m-1)\|_2^2 	 \\
&\leq \min_{{\boldsymbol w}_1}\|{\boldsymbol w}_1{\boldsymbol { Y}} - \lambda\tilde{\boldsymbol y}(0:N-m-1)\|_2^2+\min_{{\boldsymbol w}_2}\|{\boldsymbol w}_2{\boldsymbol { S}}_{ESN} - (1-\lambda)\tilde{\boldsymbol y}(0:N-m-1)\|_2^2\\
&= \lambda^2 r_{W} +(1-\lambda)^2r_{ESN},
\end{align}
where 
\begin{align}
r_{W} = \min_{{\boldsymbol w}_1}\|(1/\lambda){\boldsymbol w}_1{\boldsymbol { Y}} - \tilde{\boldsymbol y}(0:N-m-1)\|_2^2,
\end{align}
\begin{align}
r_{ESN} = \min_{{\boldsymbol w}_2}\|(1/(1-\lambda)){\boldsymbol w}_2{\boldsymbol { S}}_{ESN} - \tilde{\boldsymbol y}(0:N-m-1)\|_2^2.
\end{align}
According to the definition of STM, we have
\begin{align}
{1\over 2}MC_{WESN} &\geq \lambda^2 MC_{W} + (1-\lambda)^2MC_{ESN}\\
&\geq  \lambda^2 MC_{W} + (1-\lambda^2)MC_{ESN}.
\end{align}
Finally, the theorem is proved by substituting $\lambda^2$ as $\lambda$.

\bibliographystyle{IEEEtran}

\begin{thebibliography}{10}
	\providecommand{\url}[1]{#1}
	\csname url@samestyle\endcsname
	\providecommand{\newblock}{\relax}
	\providecommand{\bibinfo}[2]{#2}
	\providecommand{\BIBentrySTDinterwordspacing}{\spaceskip=0pt\relax}
	\providecommand{\BIBentryALTinterwordstretchfactor}{4}
	\providecommand{\BIBentryALTinterwordspacing}{\spaceskip=\fontdimen2\font plus
		\BIBentryALTinterwordstretchfactor\fontdimen3\font minus
		\fontdimen4\font\relax}
	\providecommand{\BIBforeignlanguage}[2]{{%
			\expandafter\ifx\csname l@#1\endcsname\relax
			\typeout{** WARNING: IEEEtran.bst: No hyphenation pattern has been}%
			\typeout{** loaded for the language `#1'. Using the pattern for}%
			\typeout{** the default language instead.}%
			\else
			\language=\csname l@#1\endcsname
			\fi
			#2}}
	\providecommand{\BIBdecl}{\relax}
	\BIBdecl
	
	\bibitem{LiuMIMOCom}
	L.~Liu, R.~Chen, S.~Geirhofer, K.~Sayana, Z.~Shi, and Y.~Zhou, ``Downlink
	{MIMO} in {LTE}-advanced: {SU}-{MIMO} vs. {MU}-{MIMO},'' \emph{{IEEE} Commun.
		Mag.}, vol.~50, no.~2, pp. 140--147, February 2012.
	
	\bibitem{yang2015fifty}
	S.~Yang and L.~Hanzo, ``Fifty years of {MIMO} detection: The road to
	large-scale {MIMO}s,'' \emph{{IEEE} Commun. Surveys Tuts.}, vol.~17, no.~4,
	pp. 1941--1988, 2015.
	
	\bibitem{rahmatallah2013peak}
	Y.~Rahmatallah and S.~Mohan, ``Peak-to-average power ratio reduction in {OFDM}
	systems: A survey and taxonomy,'' \emph{{IEEE} Commun. Surveys Tuts.},
	vol.~15, no.~4, pp. 1567--1592, 2013.
	
	\bibitem{chen2003iterative}
	H.~Chen and A.~M. Haimovich, ``Iterative estimation and cancellation of
	clipping noise for {OFDM} signals,'' \emph{{IEEE} Commun. Lett.}, vol.~7,
	no.~7, pp. 305--307, 2003.
	
	\bibitem{morgan2006generalized}
	D.~R. Morgan, Z.~Ma, J.~Kim, M.~G. Zierdt, and J.~Pastalan, ``A generalized
	memory polynomial model for digital predistortion of rf power amplifiers,''
	\emph{{IEEE} Trans. Signal Process.}, vol.~54, no.~10, pp. 3852--3860, 2006.
	
	\bibitem{joung2015survey}
	J.~Joung, C.~K. Ho, K.~Adachi, and S.~Sun, ``A survey on
	power-amplifier-centric techniques for spectrum- and energy-efficient
	wireless communications,'' \emph{{IEEE} Commun. Surveys Tuts.}, vol.~17,
	no.~1, pp. 315--333, First Quarter 2015.
	
	\bibitem{o2017introduction}
	T.~J. O'Shea and J.~Hoydis, ``An introduction to deep learning for the physical
	layer,'' \emph{{IEEE} Trans. on Cogn. Commun. Netw.}, vol.~3, no.~4, pp.
	563--575, 2017.
	
	\bibitem{S_Dorner}
	S.~D{ö}rner, S.~Cammerer, J.~Hoydis, and S.~ten Brink, ``Deep learning based
	communication over the air,'' \emph{{IEEE} J. Sel. Topics Signal Process.},
	vol.~12, no.~1, pp. 132--143, Feb 2018.
	
	\bibitem{farsad2018neural}
	N.~Farsad and A.~Goldsmith, ``Neural network detectors for molecular
	communication systems,'' in \emph{IEEE 19th Intl. Workshop on Signal Process.
		Adv. in Wireless Commun. (SPAWC)}, 2018, pp. 1--5.
	
	\bibitem{karanov2018end}
	B.~Karanov, M.~Chagnon, F.~Thouin, T.~A. Eriksson, H.~B{\"u}low, D.~Lavery,
	P.~Bayvel, and L.~Schmalen, ``End-to-end deep learning of optical fiber
	communications,'' \emph{J. Lightw. Technol.}, vol.~36, no.~20, pp.
	4843--4855, 2018.
	
	\bibitem{Karanov:19}
	B.~Karanov, D.~Lavery, P.~Bayvel, and L.~Schmalen, ``End-to-end optimized
	transmission over dispersive intensity-modulated channels using bidirectional
	recurrent neural networks,'' \emph{Opt. Express}, vol.~27, no.~14, pp.
	19\,650--19\,663, Jul 2019.
	
	\bibitem{khan2017machine}
	F.~N. Khan, C.~Lu, and A.~P.~T. Lau, ``Machine learning methods for optical
	communication systems,'' in \emph{Signal Process. in Photon. Commun.}\hskip
	1em plus 0.5em minus 0.4em\relax Opt. Society of America, 2017, pp. SpW2F--3.
	
	\bibitem{He2018}
	H.~Ye, G.~Y. Li, and B.~Juang, ``Power of deep learning for channel estimation
	and signal detection in {OFDM} systems,'' \emph{{IEEE} Commun. Lett.},
	vol.~7, no.~1, pp. 114--117, Feb 2018.
	
	\bibitem{zhang2019artificial}
	J.~Zhang, C.-K. Wen, S.~Jin, and G.~Y. Li, ``Artificial intelligence-aided
	receiver for a {CP}-free {OFDM} system: Design, simulation, and experimental
	test,'' \emph{arXiv preprint arXiv:1903.04766}, 2019.
	
	\bibitem{jiang2018artificial}
	P.~Jiang, T.~Wang, B.~Han, X.~Gao, J.~Zhang, C.-K. Wen, S.~Jin, and G.~Y. Li,
	``Artificial intelligence-aided {OFDM} receiver: Design and experimental
	results,'' \emph{arXiv preprint arXiv:1812.06638}, 2018.
	
	\bibitem{tan2018improving}
	X.~Tan, W.~Xu, Y.~Be'ery, Z.~Zhang, X.~You, and C.~Zhang, ``Improving massive
	{MIMO} belief propagation detector with deep neural network,'' \emph{arXiv
		preprint arXiv:1804.01002}, 2018.
	
	\bibitem{Samuel2018}
	A.~W. Neev~Samuel, Tzvi~Diskin, ``Learning to detect,'' \emph{arXiv preprint
		arXiv:1805.07631}, 2018.
	
	\bibitem{szegedy2015going}
	C.~Szegedy, W.~Liu, Y.~Jia, P.~Sermanet, S.~Reed, D.~Anguelov, D.~Erhan,
	V.~Vanhoucke, and A.~Rabinovich, ``Going deeper with convolutions,'' in
	\emph{Proc. IEEE conf. on Computer Vision and Pattern Recognition (CVPR)},
	2015, pp. 1--9.
	
	\bibitem{goodfellow2016deep}
	I.~Goodfellow, Y.~Bengio, and A.~Courville, \emph{Deep learning}.\hskip 1em
	plus 0.5em minus 0.4em\relax MIT press, 2016.
	
	\bibitem{werbos1990backpropagation}
	P.~J. Werbos \emph{et~al.}, ``Backpropagation through time: what it does and
	how to do it,'' \emph{Proc. {IEEE}}, vol.~78, no.~10, pp. 1550--1560, 1990.
	
	\bibitem{pascanu2013difficulty}
	R.~Pascanu, T.~Mikolov, and Y.~Bengio, ``On the difficulty of training
	recurrent neural networks,'' in \emph{Int. Conf. on Machine Learning (ICML)},
	2013, pp. 1310--1318.
	
	\bibitem{hochreiter1997long}
	S.~Hochreiter and J.~Schmidhuber, ``Long short-term memory,'' \emph{Neural
		Comput.}, vol.~9, no.~8, pp. 1735--1780, 1997.
	
	\bibitem{LTE_standards}
	\emph{Physical channels and modulation}, 3{GPP} Std. TS 36.211, Rev. 13.2.0,
	2016.
	
	\bibitem{jaeger2001echo}
	H.~Jaeger, ``The “echo state” approach to analysing and training recurrent
	neural networks-with an erratum note,'' \emph{Bonn, Germany: German National
		Research Center for Inf. Technol. GMD Technical Report}, vol. 148, no.~34,
	p.~13, 2001.
	
	\bibitem{mosleh2017brain}
	S.~Mosleh, L.~Liu, C.~Sahin, Y.~R. Zheng, and Y.~Yi, ``Brain-inspired wireless
	communications: Where reservoir computing meets {MIMO-OFDM},'' \emph{{IEEE}
		Trans. Neural Netw. Learn. Syst.}, vol.~29, no.~10, pp. 4694--4708, Oct 2018.
	
	\bibitem{mosleht2016energy}
	S.~Mosleh, C.~Sahin, L.~Liu, R.~Zheng, and Y.~Yi, ``An energy efficient
	decoding scheme for nonlinear {MIMO}-{OFDM} network using reservoir
	computing,'' in \emph{IEEE Int. Joint Conf. on Neural Networks (IJCNN)},
	2016, pp. 1166--1173.
	
	\bibitem{R2018}
	R.~{Shafin}, L.~{Liu}, J.~{Ashdown}, J.~{Matyjas}, M.~{Medley}, B.~{Wysocki},
	and Y.~{Yi}, ``Realizing green symbol detection via reservoir computing: An
	energy-efficiency perspective,'' in \emph{IEEE Int. Conf. on Commun. (ICC)},
	May 2018.
	
	\bibitem{barbero2008fixing}
	L.~G. Barbero and J.~S. Thompson, ``Fixing the complexity of the sphere decoder
	for {MIMO} detection,'' \emph{{IEEE} Trans. Wireless Commun.}, vol.~7, no.~6,
	2008.
	
	\bibitem{jaeger2001short}
	H.~Jaeger, \emph{Short term memory in echo state networks}.\hskip 1em plus
	0.5em minus 0.4em\relax German National Research Institute for Computer
	Science (GMD) Report, 2001, vol.~5.
	
	\bibitem{coleri2002channel}
	S.~Coleri, M.~Ergen, A.~Puri, and A.~Bahai, ``Channel estimation techniques
	based on pilot arrangement in {OFDM} systems,'' \emph{{IEEE} Trans.
		Broadcast.}, vol.~48, no.~3, pp. 223--229, 2002.
	
	\bibitem{dong2007linear}
	X.~Dong, W.-S. Lu, and A.~C. Soong, ``Linear interpolation in pilot symbol
	assisted channel estimation for ofdm,'' \emph{{IEEE} Trans. Wireless
		Commun.}, vol.~6, no.~5, 2007.
	
	\bibitem{duport2016fully}
	F.~Duport, A.~Smerieri, A.~Akrout, M.~Haelterman, and S.~Massar, ``Fully
	analogue photonic reservoir computer,'' \emph{Scientific Reports}, vol.~6, p.
	22381, 2016.
	
	\bibitem{vandoorne2014experimental}
	K.~Vandoorne, P.~Mechet, T.~Van~Vaerenbergh, M.~Fiers, G.~Morthier,
	D.~Verstraeten, B.~Schrauwen, J.~Dambre, and P.~Bienstman, ``Experimental
	demonstration of reservoir computing on a silicon photonics chip,''
	\emph{Nature Commun.}, vol.~5, p. 3541, 2014.
	
	\bibitem{chang2012algorithm}
	R.~Y. Chang, W.-H. Chung, and S.-J. Lin, ``A* algorithm inspired
	memory-efficient detection for {MIMO} systems,'' \emph{{IEEE} Wireless
		Commun. Lett.}, vol.~1, no.~5, pp. 508--511, 2012.
	
\end{thebibliography}

\end{document}